\documentclass[11pt]{article}
\usepackage{epsfig}
\usepackage{amsmath, amssymb}
\begin{document}
\newcommand{\PADS}{{\rm P}\ADS} 
\newcommand{\ads}{{\rm AdS}} \newcommand{\kg}{{\rm KG}} 
\newcommand{\fish}{{\rm fish}} 
\newcommand{\cft}{{\rm CFT}} \newcommand{\inv}{^{-1}}
\newcommand{\NN}{{\mathbb{N}}} \newcommand{\MM}{{\mathbb{M}}} 
\newcommand{\RR}{{\mathbb{R}}} \newcommand{\CC}{{\mathbb{C}}}
\newcommand{\ZZ}{{\mathbb{Z}}} \newcommand{\CM}{{\rm CM}}
\newcommand{\M}{{\mathbb{M}}}  
\newcommand{\HH}{{\mathcal H}} \newcommand{\LL}{{\mathcal L}}
\newcommand{\FF}{{\mathcal F}} \newcommand{\OO}{{\mathcal O}}
\newcommand{\supp}{{\rm supp\,}}
\newcommand{\Ad}{{\rm Ad\,}}
\renewcommand{\theequation}{\thesection.\arabic{equation}}
\def\wick#1{{\colon\!#1\colon\!}} 
\def\Li{\mathrm{Li}} \def\dilog{\mathrm{dilog}}
\def\bea{\begin{eqnarray}}\def\eea{\end{eqnarray}}
\def\ba{\begin{array}}\def\ea{\end{array}}

\title{\bf Protecting the conformal symmetry \\ via bulk renormalization
  on Anti deSitter space\footnote{Supported in part by the German
    Research Foundation (Deutsche Forschungsgemeinschaft (DFG))
    through the Institutional Strategy of the University of
    G\"ottingen, and DFG Grant RE 1208/2-1.}}
\author{Michael D\"utsch$^{1,2}$
and
Karl-Henning Rehren$^{1,2}$
}
\date{}

\maketitle

\vskip-10mm

\begin{center}
\footnotesize
$^1$ Institut f\"ur Theoretische Physik, Universit\"at G\"ottingen, \\
Friedrich-Hund-Platz 1, 37077 G\"ottingen, Germany \\[2mm]
$^2$ Courant Research Centre ``Higher Order Structures in Mathematics'',
  Universit\"at G\"ottingen, \\ Bunsenstr.\ 3--5, 37073 G\"ottingen, Germany
\end{center}

\vskip10mm

\begin{center}
{\sl Dedicated to Raymond Stora on the occasion of his 80th birthday}
\end{center}

\begin{abstract} 
The problem of perturbative breakdown of conformal symmetry can be
avoided, if a conformally covariant quantum field $\varphi$ on
$d$-dimensional Minkowski spacetime is viewed as the boundary limit of
a quantum field $\phi$ on $d+1$-dimensional anti-deSitter spacetime (AdS).  
We study the boundary limit in renormalized perturbation theory with
polynomial interactions in AdS, and point out the differences as
compared to renormalization directly on the boundary. In particular,
provided the limit exists, there is no conformal anomaly. We compute
explicitly the one-loop ``fish diagram'' on $\ads_4$ by differential
renormalization, and calculate the anomalous dimension of the
composite boundary field $\varphi^2$ with bulk interaction $\kappa
\phi^4$.    
\end{abstract}

\vskip10mm

PACS 2001: 11.10.-z, 04.62.+v


\newpage
\tableofcontents
\section{Introduction} 
\setcounter{equation}{0}
When a scale invariant free field is perturbed by an interaction, the
scaling symmetry is in general broken. In the case of the free
massless scalar field in 4-dimensional Minkowski space, this 
``conformal anomaly'' is well known: the renormalization of loop
diagrams requires the introduction of a scale parameter which breaks
scale invariance.  Using the non-uniqueness of renormalization, the
best one can reach is ``almost homogeneous scaling'', i.e.\ the
breaking terms for the scaling $x\mapsto \lambda x$ are proportional
to some power of $\log\>\lambda$. (For a systematic treatment in the
framework of causal perturbation theory see \cite{HW,DF3}.) 

In this paper, we want to address the analogous issue for scale
invariant {\em generalized free fields} (free fields with
non-canonical scaling dimension, see \eqref{weightf} below). Such
fields naturally arise as boundary limits of Klein-Gordon fields on
AdS \cite{Witten,BBMS}. The basic question is: 

\begin{itemize}
\item \it
Is it possible to construct {\em scale invariant interacting fields}
(admitting for anomalous dimensions)
\bea
(\varphi^l)_{\kappa \LL}(x)=\wick{\varphi^l(x)}+{\cal O}(\kappa)
\eea
as perturbative expansions around Wick powers $\wick{\varphi^l(x)}$ of
scale invariant generalized free fields $\varphi$ \cite{DR1}?
\end{itemize}
($L$ denotes the interaction density and $\kappa$ the coupling constant.) 

Perturbation theory around a generalized free field (in Minkowski space) 
suffers from a huge arbitrariness which is due to renormalization, as
we point out in Sect.~\ref{sec:procedure}. On the other hand, the   
requirement of scale invariance is very restrictive. In important
cases (which we do not want to exclude) it cannot be fulfilled even 
for tree diagrams (Sect.~\ref{sec:comparison}). Namely, the propagator
needs a nontrivial renormalization if the scaling dimension $\Delta$
is $\geq 2$ in four dimensions ($\frac d2$ in $d$ dimensions), and for
integer $\Delta$ a breaking of scale invariance cannot be avoided. 

We propose here a method to circumvent these difficulties and
construct perturbatively interacting fields with unbroken conformal
symmetry, by taking advantage of the AdS-CFT correspondence.
Viewing a conformally covariant field on Minkowski space-time as a
boundary limit of an AdS covariant field on 
Anti-deSitter space-time \cite{Witten,BBMS,DR}, an AdS invariant
renormalization in the bulk guarantees an anomaly free conformal
symmetry of the boundary field, provided the boundary limit exists. In
this way, the AdS-CFT correspondence turns out to be a useful tool
also when one is only interested in CFT in Minkowski spacetime.   

\medskip

In \cite{BBMS} and \cite{DR1} it was shown that the boundary limit
$z\searrow 0$ 
   \footnote{We use Poincar\'e coordinates $X\equiv(z,x^\mu)\in
   \RR_+\times \RR^d$ of $\ads_{d+1}$ such that
   $\xi=z\inv\big(x^\mu,\frac12(z^2-x^2-1),\frac12(z^2-x^2+1)\big)$
   lies on the hyperboloid $\xi\cdot\xi=1$ w.r.t.\ to the metric of
   signature $(+,-\ldots-,+)$ in the ambient space $\RR^{d+2}$. The
   $\ads$ metric is the induced one: $ds^2 = z^{-2}(dx_\mu dx^\mu -
   dz^2)$, see e.g.\ \cite{BBMS}.}  
of the scalar Klein-Gordon field $\phi(z,x)$ of mass $M$ on
$(d+1)$-dimensional AdS is a generalized free field $\varphi(x)$ with 
scaling dimension 
\bea
\label{dim}
\Delta=\Delta_+=\frac d2+\nu,\qquad \Big(\nu=\sqrt{\frac{d^2}4+M^2}\Big),
\eea
see Sect.~\ref{sec:procedure}. The corresponding boundary limit of the
free Wick powers $W(z,x)=\wick{\phi^l(z,x)}$ yields fields
$w(x)=\wick{\varphi^l(x)}$ which have scaling dimensions $l\Delta$. 
Notice that in the Witten model \cite{Witten}
of Maldacena's conjectured AdS-CFT correspondence \cite{M}, one
studies instead the ``dual'' field with boundary conditions
corresponding to $\Delta_-=\frac d2-\nu$, which is coupled to the
sources in a ``dual'' way. However, it was shown in \cite{DR} that the
dual coupling modifies the relevant bulk propagator by a correction
term in such a way, that the full propagator becomes that of the above
Klein-Gordon field, and the unrenormalized perturbative expansion of
the dually coupled boundary field is formally equivalent to the
boundary limit of the bulk field $\phi(z,x)$ with the same
interaction. (The same nontrivial features, that are of representation
theoretic nature, were established for the propagators of tensor
fields of any rank \cite{Lect}.)   

Regarding the generalized free field as a limit of a canonical free
field on AdS, the task is to extend this relation to the renormalized
interacting fields. Hence, we first construct the interacting AdS
fields 
\bea
W_{\kappa \LL}(z,x)= \wick{\phi^l(z,x)}+{\cal O}(\kappa)
\eea
for polynomial interactions $\LL=\phi^k$ in Sect.~\ref{sec:phi} and 
Sect.~\ref{sec:phi2},
using standard renormalization methods of causal perturbation theory
(reviewed in Sect.~\ref{sec:causalpt} and \ref{sec:renorm}). At
this stage, the non-uniqueness of the renormalization can be
classified by the usual short distance power counting \cite{BF,HW},
and the propagator is unique and AdS-invariant, hence the AdS symmetry
is fully preserved. 

Then, the essential step is to investigate the existence of a boundary
limit 
\bea
\label{wkl}
w_{\kappa \LL}(x) = \lim_{z\searrow 0} z^{-\Delta^W_{\kappa\LL}}
\cdot W_{\kappa \LL}(X)  
\eea 
in the renormalized theory. Here, we admit for anomalous dimensions, 
i.e., $\Delta^W_{\kappa\LL} = l\Delta + {\cal O}(\kappa)$. If this
limit exists, we prove that it inherits the AdS symmetry of the bulk
as an exact (unbroken) conformal symmetry (Sect.~\ref{sec:conf}). 

Our main result is that the boundary limit does exist, for typical
polynomial interactions, for the interacting field
(Sect.~\ref{sec:phi}) and for composite fields (Sect.~\ref{sec:phi2}),
due to nontrivial cancellations within the renormalized one-loop
distributions taking place in the limit. Although the actual
computations are ``hidden'' in Apps.~\ref{app-pf} and \ref{app:limit},
these cancellations constitute the essential mechanism to allow the
passage to the boundary.

In order to establish this result, along the way we develop a
``universal'' formula (Lemma~B.1 in App.~\ref{app:lemma}) that controls
the asymptotic behaviour near the boundary of a large class of typical
interactions and diagrams. 

\medskip

Thus, the above posed question gets an affirmative answer for those
interactions $L[\varphi(x)]$ of the conformal field which are
``induced'' by the corresponding polynomial AdS interaction
$\LL[\phi(X)]$ (as indicated by retaining the subscript $\kappa\LL$ in 
\eqref{wkl} also for the boundary field). This means \cite{DR1} that      
\bea
\kappa\int d^dx \,L[\varphi(x)] = \kappa\int
dz\,d^dx\,\sqrt{-g}\,\LL[\phi(z,x)], 
\eea
hence the CFT interaction density 
\bea
L[\varphi(x)] = \int dz\,\sqrt{-g}\, \LL[\phi(z,x)]=\int \frac{dz}{z^{d+1}}\,
\LL[\varphi_{h_z}(x)]
\label{action}
\eea
arises as the $z$-integral over $\LL[\varphi_{h_z}(x)]$ where
$\varphi_{h_z}(x)$ is the AdS field $\phi(z,x)$ re-expressed as a
family of boundary generalized free fields belonging to the Borchers
class of $\varphi$ (\cite{DR1}, see Sect.~\ref{sec:freefields}). 
We point out that, due to the integration in \eqref{action}, the
interaction vertices ``remain in the bulk''. In this sense, the
situation is converse to R\"uhl's reconstruction \cite{Ruehl} of an
AdS field from an interacting conformal field where the AdS
interaction is restricted to the boundary (namely, the AdS field in
\cite{Ruehl} satisfies the free field equation in the bulk). 

It is an essential aspect of our approach that, while the general
principles of renormalization are the same, the detailed
implementation of the rules differ in the bulk and on the boundary. 
In order to exhibit the methodic difference which allows the
renormalization in the bulk to preserve the symmetry that is
necessarily broken by renormalization on the boundary, we compare both
approaches in Sect.~\ref{sec:countex} with a flat space toy model,
where this difference is much more transparent.

\section{The general strategy}
\label{sec:procedure}
\setcounter{equation}{0} 
\subsection{Free fields}
\label{sec:freefields}
Let us recall \cite{DR1} how the Klein-Gordon field on $(d+1)$-dimensional
Anti-deSitter space and generalized free fields on $d$-dimensional
Minkowski space can be represented in terms of the same 
creation and annihilation operators, and hence as field operators on
the same Hilbert space.

The free Klein-Gordon field $\phi$ of mass $M$ on AdS can be expressed as 
\bea 
\phi(z,x) = {\textstyle \frac 1{\sqrt 2}} \; z^{\frac d2} 
\int_0^\infty dm^2 \, J_\nu(mz)\varphi_m(x)\ ,
\label{phi:kg}
\eea
where $\varphi_m$ is a massive free boundary field given by 
\bea
 \varphi_m(x)\equiv\int_{k_0\geq 0}d^dk\,\delta(k^2-m^2)
\; \big[a(k) e^{-ikx} + a^+(k)e^{ikx}\big]\,.
\label{ads:kg}
\eea 
The parameter $\nu >-1$ is related to the mass by $M^2=\nu^2-\frac{d^2}{4}$. 
The functions $z^{d/2}J_\nu(\sqrt{k^2}z)\exp\pm ikx$ are the
plane-wave solutions to the Klein-Gordon equation on AdS, where the
Laplacian is  
\bea
\label{laplace}
\square_X = -z^{1+d}\partial_z
z^{1-d}\partial_z + z^2\square_x\ ,
\eea
and $a(k),\>a^+(k)$ ($k\in\RR^d$) are creation and
annihilation operators normalized as
\bea
  [a(k),a^+(k')] = (2\pi)^{-(d-1)}\delta^d(k-k')\ ,
  \qquad [a,a] = 0 = [a^+,a^+] \ ,
\eea
in the Fock space $\HH$ over the continuous mass 1-particle space 
$\HH_1 = L^2(V_+,d^dk)$. 

In this Hilbert space, the fields
\bea
  \varphi_h(x)\equiv \int_{V_+} d^dk \; 
  h(k^2) \; \big[a(k) e^{-ikx} + a^+(k)e^{ikx}\big] 
\label{gff:h}
\eea
(with $h$ any sufficiently smooth polynomially bounded real function
on $\RR_+$) are local and Poincar\'e covariant generalized free scalar
fields in $d$-dimensional Minkowski space with K\"allen-Lehmann
measure $d\mu(m^2)=h(m^2)^2dm^2$. Thus, $\phi$ may be written as
\bea
  \phi (z,x)=\varphi_{h_z}(x)\quad\textrm{with}\quad
  h_z(m^2)\equiv {\textstyle \frac 1{\sqrt 2}} \; z^{\frac d2}
  J_\nu(zm)\ .
\label{phi=varphi_h}
\eea
Taking the boundary limit, we get \cite{BBMS,DR1}:
\bea
  \lim_{z\searrow 0} z^{-\Delta} \phi(z,x) = \varphi(x)
\label{boundlim}
\eea
with
\footnote{It should not lead to confusion that the present field 
$\varphi$ was denoted $\varphi^{(\Delta)}$ in \cite{DR1}, whereas 
$\varphi_h$ with $h(m^2)=1$ was denoted $\varphi$.}
\bea
  \varphi(x)=C_\nu\int_{V_+} d^dk \, (k^2)^{\frac{\nu}{2}}
  \, [a(k) e^{-ikx} + a^+(k)e^{ikx}]\ ,\quad\textstyle
  \Delta\equiv\nu+\frac d2\ ,\quad
  C_\nu\equiv{\frac{2^{-\nu-\frac 12}}{\Gamma(\nu+1)}}\ ,
\label{gff}
\eea
i.e., $\varphi=\varphi_h$ with $h(m^2)=C_\nu\, m^\nu$. Its
K\"allen-Lehmann measure being a homogeneous function of the mass:
\bea
  d\mu(m^2)=C_\nu^2 \;m^{2\nu}\, dm^2\ ,
\label{weightf}
\eea
the boundary field $\varphi$ is scale invariant:
\bea
  U(\lambda)\>\varphi(x)\>U(\lambda)^*=\lambda^{\Delta}
\varphi(\lambda x)\ ,
\label{scaling}
\eea
and in fact transforms like a conformal scalar field under the
representation of the AdS symmetry group on the Fock space of
the AdS Klein-Gordon field $\phi$.

\medskip

The boundary limit \eqref{boundlim} can obviously be generalized to
arbitrary Wick polynomials $W=\wick{\prod_{j=1}^l\partial_x^{a_j}\phi}$,
\bea
w(x)=\lim_{z\searrow 0} z^{-l\Delta}\;W(z,x)=\wick{ \prod_{j=1}^l
\partial^{a_j}\varphi(x)} 
\label{boundlim-W}
\eea
which have scaling dimension $D^W = l\Delta + \sum_j\vert a_j\vert$
(where $a_j\in(\NN_0)^d$ is a multi-index).

\subsection{Causal perturbation theory}
\label{sec:causalpt}
The aim of this paper is to investigate causal perturbation theory 
\cite{EG} around the generalized free field \eqref{gff} (and its Wick
polynomials \eqref{boundlim-W}). Causal perturbation theory proceeds
\cite{EG,BF,DF3} by defining, for each Wick polynomial
$W$ of free fields $\phi$, the interacting field $W_{g\LL}$ as formal
expansion in Wick products of the free field $\phi$ with distributional
coefficients. This expansion is obtained as the exponential series
of retarded products of $W$ with the interaction $g\LL$, where the
retarded products are operator-valued distributions. They are determined 
recursively (by the postulated causal properties of the interacting fields) 
{\em at non-coinciding points only}; the renormalization of the
perturbative expansion consists in the {\em extension of these
  distributions to coinciding points}. ``Renormalization conditions''
(covariance, Ward identities, \dots) serve to reduce the arbitrariness
in the extension, and the main problem is to decide whether all
desirable renormalization conditions can be fulfilled at the same
time, with a finite number of free parameters remaining. 

This program is performed with the interaction being cut off in space
and time by means of a space-time dependent coupling constant $g(x)$. 
It then remains to control the {\em adiabatic limit} of removing the
cutoff, $g(x)\to\kappa$. This limit is in general plagued by infrared
problems; it is, however, possible to define the 
{\em algebraic adiabatic limit} \cite{BF}, i.e., the local field
algebras $\FF_{\kappa\LL}(K)$ in arbitrary bounded space-time regions
$K$, without infrared problems as long as the construction of the
interacting vacuum state is postponed. 

Causal perturbation theory around a generalized free field is,
however, problematic for the following reason. To construct the
general solution for the perturbative $S$-matrix one has to
use the Wick expansion formula for time-ordered or retarded
products (also called the ``causal Wick expansion'')
\cite{EG,BF,DF3}. For simplicitly, let us discuss here the ordinary
Wick expansion formula, which for mass shell free fields is  
\bea
\label{wick}
  \wick{\varphi_m^{k_1}(x_1)}\ldots \wick{\varphi_m^{k_n}(x_n)}
= \hspace{85mm}\\ 
\sum_{r_1,\dots,r_n}\prod_{i=1}^n \Bigl(\!\begin{array}{c} k_i
  \\ r_i\end{array}\! \Bigr)\,(\Omega, \wick{\varphi_m^{k_1-r_1}(x_1)}\ldots 
\wick{\varphi_m^{k_n-r_n}(x_n)}\Omega)\cdot \wick{\varphi_m^{r_1}(x_1)\ldots 
\varphi_m^{r_n}(x_n)}\,.
\notag\eea 
For generalized free fields, the Lagrangean can be any field
relatively local w.r.t.\ the generalized free field, i.e., any element
of its Borchers class. The Borchers class contains at least the
``generalized Wick polynomials'' \cite{DR1}  
\bea
\label{gwp} 
(\wick{\varphi^l})_h(x) = \hspace{110mm} \\
 \int_{V_+} d^dk_1 \ldots\int_{V_+} d^dk_l \;
h(k_1^2,\ldots ,k_l^2)  \cdot \wick{[a(k_1) e^{-ik_1x} + \hbox{h.c.}]\ldots 
[a(k_l) e^{-ik_lx} + \hbox{h.c.}]}
\notag
\eea
where $h:(\RR_+)^{l}\rightarrow \CC$ is any symmetric and sufficiently
regular function. Let us choose a Lagrangean
$L(y)=(\wick{\varphi^4})_H(y)$ with an arbitrary function
$H(k_1^2,\dots,k_4^2)$. It is then easy to see, that the Wick
expansion of, say, $\varphi_h(x)$ with $L(y)$ does not 
factorize as in \eqref{wick}, but rather contains terms of the form
\bea\label{gwick}
 \int_{V_+}\!\! d^dk_1 \dots d^dk_3\; 
h(x-y;k_1^2,k_2^2,k_3^2)  \; \wick{[a(k_1) e^{-ik_1y} + \hbox{h.c.}]\ldots 
[a(k_3) e^{-ik_3y} + \hbox{h.c.}]}\,,\;
\eea
where 
\bea 
h(x-y;k_1^2,k_2^2,k_3^2) = \int_{V_+}dq\; e^{-iq(x-y)}\,h(q^2)\,
H(q^2,k_1^2,k_2^2,k_3^2).
\eea
Because the dependence of this function on $x-y$ and on $k_i^2$ is
entangled in a nontrivial manner, the numerical distribution cannot be
separated from the operator-valued distribution as in \eqref{wick}
(unless $H$ happens to be a factorizing function). Interpreting
\eqref{gwick} as an operator 
product expansion, reveals a characteristic feature of the theory of
generalized free fields: performing first the $k$-integrations, the
subsequent $q$-integration may be interpreted as a ``continuous sum''
over generalized Wick products. More importantly, however, the failure
of separation as in \eqref{wick} would require more refined methods to
establish the existence of a renormalization, than the standard
methods of causal perturbation theory, which proceeds by renormalizing
only the numerical distributions (see below).    

\medskip 

Let us contrast the general case to the case when the interaction is
induced by a local interaction on AdS \cite{DR1} as described in the
introduction, i.e., when the conformal field $\varphi$ arises as the
boundary limit of a canonical AdS field $\phi$ with interaction
$\kappa\LL$. The Lagrangean $L$ given by \eqref{action} with, say, 
$\LL=\phi^4$ on AdS is $L= (\wick{\varphi^4})_H$ with
\bea
H(k_1^2,\dots,k_4^2)=\int dz\,z^{-d-1}\prod h_z(k_i^2),
\eea
i.e., $H$ is a $z$-integral over factorizing functions; one can
therefore reorganize the continuous OPE as a $z$-integral over Wick
products of the distinguished fields $\varphi_{h_z}(x)$ as in
\eqref{phi=varphi_h}, rather than generalized Wick products as in
\eqref{gwp}. This fact seems to reduce the renormalization
ambiguity drastically, since the freedom is only in the choice of
suitable weight functions in $z$. Whether a conformally covariant
renormalization of the OPE of perturbed boundary fields is possible,
would require a nontrivial analysis.  

This is the reason why we propose to work instead with the ``bulk
approach'' mentioned before, using the correspondence \eqref{boundlim}
and \eqref{boundlim-W}; i.e., we first construct the perturbative
interacting fields on $(d+1)$-dimensional Anti-deSitter space
\cite{BF,HW}, and then study their boundary limit. We shall see that
conformal covariance can be maintained on the boundary because 
AdS covariance can be maintained in the bulk. The issue therefore has
been shifted to the existence of the limit. It will be illustrated in
Sect.~\ref{sec:countex}, why this indirect approach gives different
results than the direct approach perturbing generalized free fields on
the boundary.

\medskip 

In \cite{BF} and \cite{HW} perturbative interacting fields have been 
constructed on an arbitrary globally hyperbolic curved spacetime
${\cal M}$ for localized interactions $G(x)\LL(x)$, i.e.,
the interaction $\LL$ is switched on by $G\in {\cal D(M)}$. The
Anti-deSitter spacetime is not itself globally hyperbolic, but its
covering is conformally equivalent to a $\ZZ_2$ quotient of a globally
hyperbolic space-time \cite{AIS}. In this way, the lack of global
hyperbolicity can be circumvented in terms of boundary conditions ``at
infinity'' ($z=0$). 

If one wants to take the boundary limit, one obviously must not cut
off the interaction on the boundary of AdS, hence we must perform a
``partial adiabatic limit'' which puts the switching function $G(z,x)$
to be 1 for $x\in K$ (a compact region $\subset \M_d$) and $z=0$.
It can be easily seen that the conclusion of \cite{BF}, i.e., the
independence of the algebraic adiabatic limit on the details of the
switching function outside the compact region of interest, holds also
true for the partial adiabatic limit. We may therefore assume that the
switching function factorizes as 
\bea
G(z,x)=\kappa \;\gamma(z)\,g(x) ,\quad\mathrm{where}\quad 
G\vert_{ [0,a] \times K}\equiv \kappa =\>\mathrm{constant}
\label{G=cg}
\eea
with $g\vert_K\equiv 1$ and $\gamma\vert_{[0,a]}\equiv
1$ for some $a>0$. In addition $g$ and $\gamma$ are smooth, $\supp g$
is compact and the support of $\gamma(z)$ is bounded for
$z\rightarrow\infty$. Since the support of such functions $G$ are not
compact in AdS, there may in principle be IR problems associated with
the partial adiabatic limit; but our explicit calculations in Sect.~3
show that these do not appear in the relevant examples. 
The (partial) algebraic adiabatic limit does not depend on the details
of the functions $g$ and $\gamma$, provided $a$ is sufficiently large.

In practice, we proceed as follows: Given a Wick monomial $w$ in the
generalized free field $\varphi$ and its derivatives, we first replace
$\varphi(x)$ by the AdS field $\phi(z,x)$ (whose boundary limit is
$\varphi(x)$), and construct the interacting AdS field
$W_{\kappa\LL}(z,x)$ associated with the corresponding Wick monomial $W$
in $\phi$ and its derivatives. Then we define the interacting field
$w_{\kappa\LL}(x)$ in $\M_d$ as boundary limit of the interacting field
$W_{\kappa\LL}(z,x)$ on $\ads$, provided this limit exists: 
\bea
w_{\kappa\LL}(x)=\lim_{z\searrow 0}
z^{-\Delta^W_{\kappa\LL}}W_{\kappa\LL}(z,x)\,, 
\label{w=limW}
\eea
where
\bea
  \Delta^W_{\kappa\LL}=l\Delta+\sum_{n=1}^\infty\kappa^n
(\Delta^W_\LL)^{(n)} \, . 
\label{Delta^W}
\eea
The deformation $l\Delta\mapsto\Delta^W_{\kappa\LL}$ (i.e., the
sequence of coefficients $(\Delta^W_\LL)^{(n)}\in\CC$, $n\geq 1$)
is determined by the requirement that the limit \eqref{w=limW}
exists.

\medskip

{\bf Remark:} In ordinary perturbative QFT the anomalous
dimension is the deviation of the scaling dimension of an
(interacting) quantum field $A_{\kappa\LL}$ from the scaling
dimension of the corresponding (interacting) classical field. 
For generalized free fields there is no obvious classical counterpart. 
Instead, we call ``anomalous dimension of $w_{\kappa\LL}$'' the
deformation of the scaling dimension {\em due to the interaction}.
In contrast to ordinary perturbative QFT, it does not come from
the breaking of scale invariance in the renormalization of loop
diagrams (we maintain the AdS-symmetry in the renormalization). Instead 
its appearance is enforced by the existence of the boundary limit.

\medskip

In causal perturbation theory on AdS, $W_{\kappa\LL}$ is given by 
\cite{BF,HW}
\bea
\label{W_GL}
  W_{\kappa\LL}(X)=\sum_{n=0}^\infty\frac{\kappa^n}{n!}
\Bigl(\prod_{r=1}^n\int_0^\infty \frac{dz_r}{z_r^{d+1}}
\gamma(z_r) \int d^dx_r\,g(x_r)\Bigr) \cdot \qquad
\\
\cdot R_{n,1}(\LL(X_1),\ldots,\LL(X_n);W(X))\ \notag 
\eea 
where $X\equiv(z,x)$ and $X_j\equiv(z_j,x_j)$. The unrenormalized
retarded products $R_{n,1}$ are determined as distributions at
non-coinciding points $X_i\neq X_j \neq X$. The result is \cite{K,DF2}
\bea
\label{R:unren}
  R_{n,1}(\LL(X_1),\ldots \LL(X_n);W(X)) = \hspace{80mm}\\ 
 (-i)^n n! {\cal S}\Big[\theta\big(x^0>x^0_{n}>x_{n-1}>\dots> x^0_{1}\big)
\cdot [\LL(X_1),[\LL(X_2)\ldots [\LL(X_n),W(X) ]\ldots ]]\Big] 
\notag
\eea 
where ${\cal S}$ means symmetrization in $X_1,\ldots,X_n$.

Now let $W= \wick{\prod_{j=1}^l\partial_x^{a_j}\phi}$ and 
$\LL=\wick{\phi^k}$. Then, using \eqref{phi=varphi_h} and \eqref{phi:kg}, 
the retarded product \eqref{R:unren} may be rewritten as 
\bea\label{R(varphi_h)}
&& R_{n,1}\bigl(\wick{ \varphi_{h_{z_1}}^{\> k}(x_1)} ,\ldots,
\wick{ \varphi_{h_{z_n}}^{\> k}
(x_n)} ;\wick{
\prod_{j=1}^l\partial_x^{a_j}\varphi_{h_z}(x)}\bigr)=\qquad\qquad\qquad 
\\
&&\qquad =(-i)^n n! {\cal S} \; 
\theta\big(x^0>x^0_{n}>x_{n-1}>\dots> x^0_{1}\big) \cdot \notag\\
&&\qquad\quad\cdot\frac{(z_1^k\ldots z_n^kz^l)^{d/2}}{2^{(l+nk)/2}}
\int\prod_{r=1}^n\prod_{s=1}^kdm^2_{rs}\,J_\nu(m_{rs}z_r)
\int\prod_{j=1}^l dm_j^2\,J_\nu(m_jz) \cdot \notag\\
&&\qquad\qquad\cdot [\wick{ \prod_s\varphi_{m_{1s}}(x_1)} ,\ldots
[\wick{ \prod_s\varphi_{m_{ns}}(x_n)} ,
\wick{ \prod_{j}\partial^{a_j}\varphi_{m_j}(x)} ]
\ldots]\ .\notag
\eea 
We emphasize that writing \eqref{R:unren} as in the left-hand side of
\eqref{R(varphi_h)} is misleading: {\it It is not a retarded product in
  Minkowski space}, but in $\ads$, defined with respect to 
the causal structure in $\ads$. In particular, the problem with the
causal Wick expansion for generalized free fields mentioned before, is
absent, and its correct {\em definition} is the right-hand side of
\eqref{R(varphi_h)}. Moreover, renormalization is needed for coinciding
AdS points $X_i=X$ only, and not on the whole submanifold $x_i=x$,
as will be discussed in the next subsection.  

In the sequel, we shall be mainly concerned with special cases of the
type
\bea R_{1,1}(\wick{\phi^k(X_1)};\phi(X)) = k\;
i\Delta(X;X_1)\theta(x^0-x^0_1)\cdot \wick{\phi(X_1)^{k-1}}
\eea  
and 
\bea
\label{typical} 
 R_{1,1}(\wick{\phi^k(X_1)};\wick{\phi^2(X)}) = 2k\
i\Delta(X;X_1)\theta(x^0-x^0_1)\cdot\wick{\phi(X_1)^{k-1}\phi(X)} +
\qquad \\ \qquad\qquad + \;
k(k-1) \; i (\Delta^+(X;X_1)^2-\Delta^+(X_1;X)^2)\theta(x^0-x^0_1)
\cdot \wick{\phi(X_1)^{k-2}}  
\notag
\eea 
where $\Delta^+(X;X_1) = (\Omega,\phi(X)\phi(X_1)\Omega)$ is the scalar
2-point function, and $\Delta(X;X_1) =
(\Omega,[\phi(X),\phi(X_1)]\Omega)$ the commutator function.

\subsection{The problem of renormalization}
\label{sec:renorm}
The expressions \eqref{R:unren}--\eqref{typical} are not defined
as distributions at coinciding points, due to the time-ordering
$\theta$ functions. The problem of renormalization is thus the
extension of the retarded products to distributions $R_{n,1}(\ldots)$
on $(\RR_+\times\RR^d)^{n+1}$. By the recursive construction principle
underlying causal perturbation theory, once this has been achieved
for $R_{l,1}$ ($l<n$), then $R_{n,1}$ is already determined everywhere
outside the {\em total} diagonal   
\bea
\Delta_{n+1}\equiv \{(X_1,...,X_n;X)\>|\> X_j=X\>\forall j=1,...,n\}\,.
\label{diagonal}
\eea
Renormalization at $n$th order is thus reduced to the extension
of the distributions $R_{n,1}$ from $(\RR_+\times\RR^d)^{n+1} \setminus
\Delta_{n+1}$ to $(\RR_+\times\RR^d)^{n+1}$.

Applying the recursion as indicated, gives rise to a diagrammatic
expansion of $R_{n,1}$ in terms of Wick products with propagators and
numerical distributions $r_{m,1}^\circ(\ldots)$ ($m\leq n$) as
coefficients, as in \eqref{typical}. The latter are the vacuum
expectation values of operator-valued distributions (with field
arguments of possibly lower order). E.g., for $W=\wick{\phi^2}$ and 
$\LL=\wick{\phi^k}$, there arises the ``fish diagram'' (Fig.~1) as
the coefficient of $\wick{\phi^{k-2}(X_1)}$ to first order in $\kappa$.  

\medskip

\indent\indent\begin{minipage}{120mm}
\hskip30mm \epsfig{file=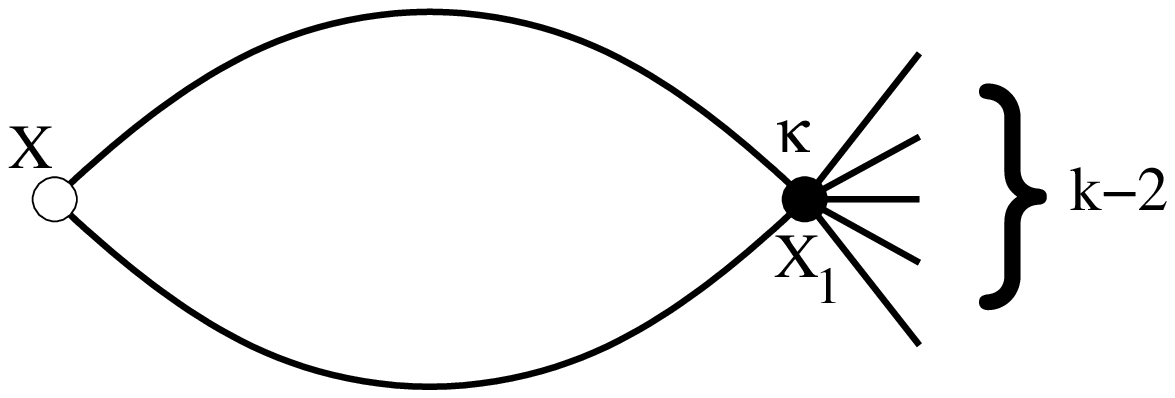,width=5.0cm}

\small{\bf Fig.~1: The ``fish'' diagram} arising in first order
perturbation theory for the interacting field
$\big(\phi^2\big)_{\kappa\LL}(X)$ with interaction $\LL=\phi^k$.
The diagram symbolizes the distribution $r_\fish(X_1;X)\equiv
(\Omega,R_{1,1}(\wick{\phi^2(X_1)},\wick{\phi^2(X)})\Omega)$ or the corresponding
unrenormalized expression $r_\fish^\circ$ (given by \eqref{5-4:r^0commu} or
\eqref{fish:nonren} resp., appearing in the second line of \eqref{typical}). 
\end{minipage}

\medskip

Renormalization is done in terms of the {\em numerical} distributions 
$r^\circ_{m,1}$, by extending them to distributions $r_{m,1}$ on
$(\RR_+\times \RR^d)^{m+1}$. 
(For an example in flat space, see Sect.~\ref{sec:countex} ) We shall
see, however, that the $z\searrow 0$ behaviour of the  renormalized
operator-valued distributions $R_{n,1}$ on AdS is in general not the
same as that of the numerical distributions $r_{m,1}$; thus the
existence of the limit has to be studied for the operator-valued
distribution $R_{n,1}$.   

For a rigorous and complete definition of the retarded products
$R_{n,1}$ we refer to the renormalization axioms given in
\cite{DF3}
\footnote{The ``off-shell'' formalism in \cite{DF3} is
  advantageous only when derivatives of fields appear as arguments of
  retarded products. In the present study, the field operators may be
  regarded as ``on-shell'', i.e., the unperturbed field satisfies the
  free equation of motion. To simplify the notation, we will, however, 
  write $W_{\kappa\LL}=(\phi^l)_{\kappa\phi^k}$ when $W=\wick{\phi^l}$
  and $\LL=\wick{\phi^k}$.}, 
with appropriate modifications due to the curvature of $\ads$ \cite{BF,HW}. 
In particular, the renormalization should not increase the scaling
degree of a distribution \cite{BF}, which controls the ``strength of
the UV singularity'': The scaling degree is defined in flat space by 
\bea
  {\rm sd}_{Y}(f(\,\cdot\,;X)) = {\rm inf} \{\delta\in\RR\vert
  \lim_{\lambda\searrow 0}\lambda^\delta f(X+\lambda Y;X)=0\}\ , 
\eea
where the limit is meant as a distribution in $Y\in\RR^{d+1}$; in
curved spacetime, $Y$ is taken in the tangent space and the argument
$X+\lambda Y$ has to be replaced by the geodesic exponential 
$\exp_X(\lambda Y)$.  

Moreover, the renormalization conditions of translation
invariance and $\LL_+^\uparrow$-covariance are replaced by
AdS-invariance (group $SO(2,d)$). The expression \eqref{R:unren}
is obviously AdS-invariant, so the problem consists in preservation of
this symmetry upon renormalization. 

Since we construct the interacting field on AdS, 
$R_{n,1}(\LL(X_1),\dots,\LL(X_n);W(X))$ needs to be renormalized
at $X_k=X$ $\forall\; k$ only, while at $x_k=x$ for all $k$,
$z_k\neq z$ for some $k$, it is already defined by the recursion. This fact
is responsible for a drastic reduction of renormalization ambiguities
in the AdS approach, as compared to renormalization of generalized
free fields on Minkowski space. 

The renormalization freedom is further reduced by {\em requiring} 
the existence of a boundary limit as a renormalization condition. We 
shall see in some typical examples (Sect.~\ref{sec:phi} and
Sect.~\ref{sec:phi2}) that this condition may require a ``field
mixing'', i.e., perturbative corrections of an interacting Wick
monomial by $\OO(\kappa)$ times other Wick monomials, in order to
cancel perturbative contributions of different scaling dimensions. 

We shall show in the next subsection that for $W=\wick{\phi^l}$ (no
derivatives), the AdS covariant renormalization of $W_{\kappa\LL}$
ensures conformal covariance of its boundary limit \eqref{w=limW}
\bea
w_{\kappa\LL}(x) = \lim_{z\searrow0} 
z^{-\Delta^W_{\kappa\LL}}\cdot W_{\kappa\LL}(z,x) 
\label{boundarylimit}
\eea
provided this limit exists, with a suitable (coupling dependent) scale
dimension $\Delta^W_{\kappa\LL}=l\Delta+\OO(\kappa)$. 

Then we shall illustrate the difference between renormalization on AdS
and renormalization on the Minkowski boundary by a flat space model
which avoids the technical complications of the curvature. 

In Sect.~3, we shall address the renormalizability on AdS and the
existence of the boundary limit \eqref{boundarylimit} with some case studies.

\subsection{Conformal symmetry}
\label{sec:conf}

In this subsection we assume that for a polynomial interaction
$\LL(\phi)$, and for $W=$ $\wick{\prod_{j=1}^l\partial_x^{a_j}\phi}$ a
Wick polynomial of the free field, an AdS-invariant renormalization 
of the interacting field $W_{\kappa\LL}$ has been achieved, and that
the boundary limit \eqref{w=limW} of $W_{\kappa\LL}$ exists with a
suitable deformation $l\Delta\mapsto\Delta^W_{\kappa\LL}$ of the power
of $z$ as in \eqref{Delta^W}. Under these assumptions we shall prove:

\medskip

\noindent {\bf Proposition 2.1:} {\sl If the boundary limit
  \eqref{w=limW} $w_{\kappa\LL}$ of $W_{\kappa\LL}$ exists, then it 
  is a scale covariant field with scaling dimension 
\bea
  D^w_{\kappa\LL}=\sum_{j=1}^l |a_j|+ \Delta^W_{\kappa\LL}\,.
\eea
If $W=\wick{\phi^l}$ contains no derivatives, then $w_{\kappa\LL}$
is a conformally covariant scalar field.}

\medskip

This is, of course, a variant of the central result in
\cite{BBMS}, that the boundary limit of a scalar AdS field, if it
exists, automatically inherits unbroken conformal symmetry. The proof
given there describes the CFT as a ``theory \`a la L\"uscher--Mack''
\cite[Sect.~3]{BBMS} on the cone $\mathrm {C}_{2,d}=\{\xi\in \RR^{d+2}:
\xi\cdot\xi=0\}$, or a covering thereof. We want to include here a
proof that refers directly to the CFT on $d$-dimensional Minkowski
spacetime $\MM_d$, which is (a chart of) the {\em projective} cone 
$\mathrm{PC}_{2,d}=\mathrm {C}_{2,d}\big/\{\xi \sim \lambda\xi\}$, or
a covering thereof. ($\mathrm{PC}_{2,d}$ is also known as the Dirac
manifold $\CM_d$.)

{\it Proof of Prop.~2.1:}
Let $U$ be the unitary representation of $SO(2,d)$ on the Fock space of 
the free Klein-Gordon field $\phi$ on AdS, which implements also the
conformal transformation of the boundary generalized free field
\cite{DR1}. For the subgroup corresponding to conformal scale
transformations on the boundary, we have 
\bea
 \Ad U(\lambda)\>\phi(z,x) \equiv U(\lambda)\,\phi(z,x)\,U(\lambda)^*
=\phi (\lambda z,\lambda x) 
\label{AdS-trans}
\eea
and hence
\bea
  \Ad U(\lambda)\>W(X)=\lambda^{\sum_j |a_j|} \, W(\lambda X)
\eea
for $W=\wick{\prod_{j=1}^l\partial_x^{a_j}\phi}$. By means of 
\eqref{R:unren} and \eqref{AdS-trans} we conclude
\bea
  \Ad U(\lambda)\>R_{n,1}(\LL(X_1),\ldots;W(X))=
\lambda^{\sum_j |a_j|} \, R_{n,1}(\LL(\lambda X),\ldots;W(\lambda X))
\eea
at non-coinciding points (using here that the interaction $\LL$
contains no derivatives of $\phi$). Since we assume that an AdS-invariant
renormalization has been achieved
\footnote{Concrete AdS-invariant renormalization schemes will be
  presented below.}, 
this identity is maintained in the extension to coinciding points. In
terms of the interacting fields \eqref{W_GL}, this gives   
\bea
  \Ad U(\lambda)\>W_{\kappa\LL}(X)=\lambda^{\sum_j |a_j|} \,
W_{\kappa\LL}(\lambda X)
\eea
in the algebraic adiabatic limit. With that and \eqref{w=limW} we obtain
\bea
\label{D=Delta:proof}
&& \Ad  U(\lambda)\> w_{\kappa\LL}(x)=\lim_{z\searrow 0}z^{-\Delta^W_{\kappa\LL}} 
\Ad U(\lambda)\> W_{\kappa\LL}(z,x) = \\ &&\qquad
=\lambda^{\sum_j |a_j|+\Delta^W_{\kappa\LL}} \,
\lim_{z\searrow 0} (\lambda z)^{-\Delta^W_{\kappa\LL}}\;
W_{\kappa\LL}(\lambda x,\lambda z)
=\lambda^{\sum_j |a_j|+\Delta^W_{\kappa\LL}} \,
w_{\kappa\LL}(\lambda x)\ .\quad\notag
\eea 
This proves the first assertion of the proposition.

\medskip

We are now going to investigate whether the conclusion 
\eqref{D=Delta:proof} applies to arbitrary AdS-transformations. Let
$t\in SO(2,d): (z,x) \mapsto (z',x')$ be an AdS-transformation, $\bar t$ 
the conformal transformation induced by $t$ on the boundary, i.e., 
$\lim_{z\searrow 0} x'(z,x)=\bar t x$. For free Wick powers 
$W=\wick{\phi^l}$ (without derivatives) and consequently 
$w=\wick{\varphi^l}$ we obtain: 
\bea
\label{wick:conf}
 \Ad U(\bar t)\> w(x)=\lim_{z\searrow 0} z^{-l\Delta}
\Ad U(t)\>\;W(z,x) = \qquad\qquad\qquad\qquad\quad
\\
=\lim_{z\searrow 0}\Bigl(\frac{z'}{z}\Bigr)^{l\Delta}
{(z')^{-l\Delta}} W(z',x') 
=\lim_{z\searrow 0}\Bigl(\frac{z'}{z}\Bigr)^{l\Delta}
w(\bar t x)\ ,
\notag \eea 
(This argument would fail if $W$ involved derivatives.) Now, 
AdS-invariance of the volume element $z^{-d-1}\,dz\,d^dx$ implies
\bea
  z^{-d-1} = z'^{-d-1}
\Big{\vert}\frac{\partial (z',x')}{\partial (z,x)}\Big{\vert}\,
\label{vol:invariant}
\eea
from which it is an easy exercise to conclude that in the limit
$z\searrow0$ (where $\lim_{z\searrow 0}\frac{\partial z'}{\partial x}=0$
and $\lim_{z\searrow 0}\frac{\partial z'}{\partial z}=\lim_{z\searrow 0}
\frac{z'}{z}$) one obtains
\bea
\lim_{z\searrow 0} \frac{z'}{z}= 
\Big{\vert}\frac{\partial (\bar t x)}{\partial x}\Big{\vert}^{1/d}.
\eea
Thus, the factor in \eqref{wick:conf} equals the conformal prefactor for
a covariant field of scaling dimension $l\Delta$.

Turning to interacting fields $W_{\kappa\LL}$ for $W=\wick{\phi^l}$,
the AdS-invariance of the retarded products, 
\bea
\Ad U(t)\>R_{n,1}(\LL(X_1),\ldots;W(X))= R_{n,1}(\LL(tX_1), \ldots;W(tX))
\label{R:AdS}
\eea 
for $t\in SO(2,d)$, implies AdS-invariance of the interacting bulk
fields in the algebraic adiabatic limit
\footnote{For a special conformal transformation $\bar t$
the function $G(t^{-1}(z,x))$ does not factorize as \eqref{G=cg}
if $G$ does; but this does not obstruct our procedure thanks to
Prop.~8.1 in \cite{BF}: in the algebraic adiabatic limit only the
constancy of $G$ in the region of interest matters, and this is
preserved by the transformation $t$.}:
\bea
 \Ad U(t)\>W_{\kappa\LL}(X) = W_{\kappa\LL}(tX)\ .
\eea
With that we find as before that $w_{\kappa\LL}(x)$ is conformally
covariant with scaling dimension $\Delta_{\kappa\LL}^W$
(provided it exists). Namely, 
\bea  \Ad U(\bar t)\>w_{\kappa\LL}(x)  =\lim_{z\searrow 0} z^{-\Delta_{\kappa\LL}^W}
\cdot \Ad U(t)\> W_{\kappa\LL}(z,x) =\qquad\qquad\qquad
\\
=\lim_{z\searrow 0}\Bigl(\frac{z'}{z}\Bigr)^{\Delta^W_{\kappa\LL}}(z')^{-\Delta^W_{\kappa\LL}}\cdot 
W_{\kappa\LL} (z',x')
=
\Big{\vert}\frac{\partial (\bar t x)}{\partial x}
\Big{\vert}^{\Delta_{\kappa\LL}^W/d}\cdot w_{\kappa\LL}(\bar t x)\ .
\notag\eea 
This completes the proof of Prop.~2.1.

\subsection{Renormalization on a submanifold: A pedagogical example}
\label{sec:countex}
We want to illustrate by a simple model that renormalization of a
field in $d+1$ dimensions and subsequent restriction to a
$d$-dimensional submanifold is not equivalent to renormalization of
the restricted fields.

Instead of $\CM_d$ as boundary of $\ads_{d+1}$, we study the
$4$-dimensional Minkowski space $\M_4$ (with coordinates 
$x=(x^\mu)_{\mu=0,...,3}\in\RR^4$ and relative coordinates $y$) as a
submanifold of the $5$-dimensional Minkowski space $\M_5$ (with
coordinates $X=(z\equiv x^4,x)\in\RR\times\M_4$ and relative
coordinates $Y=(u,y)$). The boundary limit \eqref{boundlim}
corresponds to the restriction to $\M_4$ of the fields in $\M_5$. 
The two-point function of a Klein-Gordon field of mass $M\geq 0$ in
$\M_d$ is given by 
\bea
  \Delta^{+(d)}_M(y)\equiv (\Omega,\phi(x+y)\phi(x)\Omega)=\frac{1}{(2\pi)^{d-1}}\int d^dp\,\theta(p^0)
\delta(p^2-M^2)e^{-ipy} 
\label{Delta^+}
\eea
Replacing $d=5$ and replacing $y$ by $Y=(u,y)$, this can be viewed as
the 2-point function of a generalized free field in $\M_4$ with
$u$-dependent K\"allen-Lehmann weight \cite{BBMS}:
\bea
\Delta^{+(5)}_M(Y)
=\frac{1}{2\pi} \int_{M^2}^\infty dm^2\;\frac{\cos(\sqrt{m^2-M^2}u)}
{\sqrt{m^2-M^2}}\;\Delta^{+(4)}_m(y)\,.
\label{5-4:2-point}
\eea 

For later reference, we also introduce the corresponding commutator
functions 
\bea
  \Delta^{(d)}_M(y)\equiv\Delta^{+(d)}_M(y)-\Delta^{+(d)}_M(-y)
\label{D_m^d}
\eea
and the retarded propagators 
\bea
  \Delta_M^{\mathrm{ret}(d)}(y)\equiv\Delta^{(d)}_M(y)\theta (y^0)
=\frac{i}{(2\pi)^d}\int d^dp\,\frac{e^{-ipy}}{p^2+ip^00-M^2}\ ,
\label{D_m^ret-d}
\eea
such that $\Delta^{(d)}_M(y)\theta (-y^0)= -\Delta_M^{\mathrm{ret}(d)}(-y)$. 

\medskip

We first investigate the renormalization of the fish diagram (Fig.~1)
in $\M_5$. This means that we have to extend the distribution
\bea
\label{5-4:r^0commu}
  r_\fish^\circ(Y)\equiv-i(\Omega,[\wick{\phi^2(X+Y)} ,
\wick{ \phi^2(X)} ]\Omega) \theta (-y^0) = \qquad\qquad \\
= -2i\left(\Delta^{+(5)}_M(Y)^2 - \Delta^{+(5)}_M(-Y)^2\right)\theta(-y^0) ,
\notag 
\eea 
which is well defined for $Y\equiv(u,y)\not= 0$ (because
$[\wick{ \phi^2(\cdot)} ,\wick{ \phi^2(\cdot)} ]$ 
vanishes for $y^2<u^2$), to a distribution $r_\fish\in {\cal D}'(\M_5)$
(i.e., to $Y= 0$). The extension has to be such that it does not
increase the scaling degree with respect to $Y\to 0$. 

To obtain a solution of the extension problem in $\M_5$, we work with
the K\"allen-Lehmann representation in $\M_5$. The square of the
2-point function is given in App.~\ref{kallen}. Choosing for
simplicity the field to be massless, this gives (using \eqref{KL} 
with $d=5$ and $m_1=m_2=0$) 
\bea
  r_\fish^\circ(Y)=\frac{|S_3|}{8(2\pi)^4}\int_0^\infty dm^2\, 
m\>i\Delta_m^{\mathrm{ret}(5)}(-Y).
\label{Mink-5:r^0}
\eea 
The UV divergence of the unrenormalized distribution $r_\fish^\circ$ shows
up in the divergence of the mass integral. The most general $SO(1,4)$ Lorentz
invariant extension with the required scaling degree is given by \cite{DF3} 
\bea\label{flatrenorm}
  r_\fish^{(\mu)}(Y)\propto (-\square_{Y}+\mu^2)\int dm^2\,
 \frac{m}{m^2+\mu^2}\;i\Delta_m^{\mathrm{ret}(5)}(-Y)\qquad (\mu^2\geq0)
\eea 
depending on a renormalization parameter $\mu$. (The symbol $\propto$
stands for suppressed numerical factors). 

We have obtained $r_\fish^{(\mu)}$ by renormalizing in $\M_5$. 
We now consider how this distribution would appear when
regarded as a distribution on the hypersurface $\M_4$ with the
transverse difference coordinate $u$ as a parameter. Writing $Y=(u,y)$
and the five-momentum as $(v,p)$, we arrive at  
\bea
r_\fish^{(\mu)}(u,y) \propto (-\square_{Y}+\mu^2)\int dm^2\;
 \frac{m}{m^2+\mu^2}\int d^4p\; e^{ipy}\int dv\;
\frac{e^{-ivu}}{m^2+v^2-p^2-ip^00}\notag\\
\propto (-\square_{y}+\partial_{u}^2+\mu^2)
\int d^4p\, e^{ipy}\int dm^2\,\frac{m}{m^2+\mu^2}
\frac{e^{-|u|\sqrt{m^2-p^2-ip^00}}}{\sqrt{m^2-p^2-ip^00}}
\quad
\label{holo-flatrenorm}\eea 
The appearance of the derivative $\partial^2_{u}$ (outside of the
integrals) is characteristic for the $5$-dimensional renormalization. 
One cannot get rid of this operator, because it cannot be shifted
under the integral. (The integrand is not differentiable with respect
to $u$ at $u=0$). It is the reason why $5$-dimensional renormalization
``as seen from the hypersurface'' goes beyond standard $4$-dimensional
renormalization. One way to understand this fact is that on the
hypersurface, the fields $\partial_z^n\phi(z,x)\vert_{z=0}$ are
independent fields which ``mix'' with $\phi\vert_{z=0}$ upon
5-dimensional renormalization.  

In order to exhibit this more clearly, we compare the result of
renormalization in the bulk with the alternative procedure of
renormalization on the hypersurface, where we have a $z$-dependent
family of fields in four dimensions, similar as in
\eqref{phi=varphi_h}. The label $z$ just distinguishes 
different generalized free fields $\varphi_z(x)\equiv\phi(x,z)$ on the same
hypersurface, see \cite{DR1}. That is, we write the $5$-dimensional
2-point functions in the unrenormalized distribution $r_\fish^\circ$
\eqref{5-4:r^0commu} as a $u=z_1-z_2$-dependent integral over
$4$-dimensional 2-point functions as in \eqref{5-4:2-point} (with
$M=0$), and apply the K\"allen-Lehmann representation for the resulting 
products of 2-point functions as in \eqref{KL} with $d=4$. This gives
\bea
  r_\fish^\circ(Y)=-\frac{2i}{(2\pi)^2}\int_0^\infty dm_1^2\int_0^\infty
dm_2^2\,\frac{\mathrm{cos}\>m_1u}{m_1}\frac{\mathrm{cos}\>m_2u}{m_2} \cdot
\qquad\qquad\qquad
\label{5-4:r^0}\\ 
\cdot\Bigl(\Delta^{+(4)}_{m_1}(y)\Delta^{+(4)}_{m_2}(y)-
\Delta^{+(4)}_{m_1}(-y)\Delta^{+(4)}_{m_2}(-y)\Bigr)\theta (-y^0) =
\notag\\ \qquad\qquad\qquad\qquad
=\int_0^\infty dm^2\, F(m^2,u)\;i \Delta_m^{\mathrm{ret}(4)}(-y)\ ,
\notag
\eea 
with
\bea
  F(m^2,u)\equiv\frac{2m^{-2}}{(2\pi)^4}\int_0^\infty dm_1\int_0^\infty
dm_2\,\theta(m-m_1-m_2)\cdot\qquad\qquad\qquad
\label{5-4:F} \\
\cdot\cos(m_1u)\>\cos(m_2u)\sqrt{
(m^2-{m_1^2}-{m_2^2})^2-4{m_1^2m_2^2}}\ .
\notag
\eea 
The unrenormalized distribution $r_\fish^\circ$ exists in 
${\cal D}'(\M_5\setminus\{0\})$, but for $Y=0$, the mass integral on
the right hand side of \eqref{5-4:r^0} diverges in the region
$m^2\rightarrow \infty$. Renormalization on $\M_4$ means regarding
\eqref{5-4:r^0} as a $u$-dependent K\"allen-Lehmann representation in
$\M_4$ and extending it to the diagonal of $\M_4$ in an $SO(1,3)$
Lorentz invariant way. 

At $u\neq0$, an extension to $y=0$ is in fact trivial because
\eqref{5-4:r^0} is already defined there, but the extension is
non-unique ($\delta$-functions in $y$). In order to extend also to
$u=0$ ($u=0$ corresponds to two fields on the same hypersurface), one
have to consider the most general $SO(1,3)$ Lorentz invariant
4-dimensional renormalization  
\bea
\label{submani}
  \tilde r_\fish^{(\mu)}
(Y):= (-\square_{y}+\mu^2)\int dm^2\,
\frac{F(m^2,u)}{m^2+\mu^2}\cdot i\Delta_m^{\mathrm{ret}(4)}(-y)\quad
\quad(\mu^2\geq 0)\ .
\eea 
These distributions exist even in ${\cal D}'(\M_5)$, have scaling degree
$\mathrm{sd}\>(\tilde r_\fish^{(\mu)})=6=\mathrm{sd}\>(r_\fish^\circ )$ and 
agree with $r_\fish^\circ$ for $y\not= 0$. So, $\tilde r_\fish^{(\mu)}(u,y)$ 
solves the renormalization (i.e.\ extension) problem in $\M_4$.  
But it is not a renormalization in $\M_5$ because it does not agree
with $r_\fish^\circ$ at $y=0\,\wedge\, u\not= 0$. To see this,
we evaluate both \eqref{5-4:r^0} and \eqref{submani} on a test function 
$G(Y)=\gamma(u)g(y)$ with $0\not\in\supp\gamma$. Suppressing irrelevant
constants, the difference is  
\bea
\tilde r_\fish^{(\mu)}(G)-r_\fish^\circ(G)\propto 
\int dm^2 \int du \,\gamma(u) F(m^2,u)\int d^4k \,\hat g(k)
\Big(\frac{k^2+\mu^2}{(m^2+\mu^2)(k^2-m^2)}-\frac{1}{k^2-m^2}\Big)
\notag \\
\label{difference}
\propto \int dm^2 \int du \,\gamma(u) \frac{F(m^2,u)}{m^2+\mu^2} \int d^4k
\,\hat g(k) \propto g(0) \int \frac{dm^2}{m^2+\mu^2} \int du \,\gamma(u)
F(m^2,u).\qquad \qquad 
\eea 
One can actually compute $F(m^2,u) = m^2 f(mu)$ by using variables
$m_1u+m_2u=mx$ and $m_1u-m_2u=my$ in \eqref{5-4:F}, giving $f(t)
\propto J_0(t)+J_2(t) = 2t\inv J_1(t)$. Thus, since $0\not\in\supp\gamma$,
the $u$-integral in \eqref{difference} decays $\sim m^{-\frac 12}$ due
to the oscillatory behaviour of $J_1$, so that the $m^2$-integral is
finite as required for a 4-dimensional renormalization. But it 
obviously does not vanish for generic $\gamma$, as would be required by a
5-dimensional renormalization. This proves the claim. Note that the
scale-invariant choice $\mu^2=0$ does not alter the conclusion (the 
mass integral in \eqref{difference} in this case is $\propto 
\int_0^\infty J_1(t)dt=1$). An
analogous but more refined argument shows, that also when one admits a
function $\mu(u)$, the resulting distribution cannot coincide with
$r_\fish^\circ$ for all $(y=0,u\neq0)$.  

The fact that renormalization performed on a submanifold (eq.\
\eqref{submani}) does not coincide with proper renormalization in the
bulk (eqs.\ \eqref{flatrenorm}, \eqref{holo-flatrenorm}), is the main
message of this subsection. The breakdown of
the bulk symmetry in the hypersurface renormalization is the
counterpart of conformal symmetry breaking in AdS-CFT. It can be
avoided by bulk renormalization, and subsequent restriction (boundary
limit).  

\section{Case studies I: The interacting boundary field
  $\varphi_{\kappa\phi^k}$}  
\label{sec:phi}
\setcounter{equation}{0} 
We proceed with some case studies concerning the compatibility of an
AdS-invariant renormalization with the existence of the boundary limit. 
We shall not endeavour the greatest possible generality;
e.g., we shall always assume the AdS mass parameter $M^2$ to be
sufficiently large to avoid the Breitenlohner-Freedman critical
behaviour in the range $\nu^2 \equiv \frac {d^2}4+M^2<1$ (see, e.g.,
\cite{BBMS}). 

We start with the perturbative construction of the interacting field
$\varphi_{\kappa\LL}$ with interaction $\LL = \wick{\phi^k}$ as a
deformation of $\varphi$. The renormalization of
$R_{1,1}(\LL(X_1),\phi(X))$ in this case is unproblematic,
but it 
serves to illustrate the difference between various approaches. 
In order to work out the boundary limit of the renormalized
bulk field $\phi_{\kappa\LL}$, we introduce a general technique of 
computation (Sect.~\ref{sec:massshift} to be used in more general
cases as well. In the subsequent section, we shall choose to study the
renormalization and boundary limit of the field $(\phi^2)_{\kappa\LL}$
because in this case, the perturbative expansion involves a loop
diagram (the fish diagram, Fig.~1) already at first order.  

\medskip

Our strategy is to construct the interacting AdS field
$\phi_{\kappa\phi^k}(X)$, and then take its boundary limit.
In the diagrammatic expansion of $\phi_{\kappa\phi^k}(X)$, each diagram
has a single propagator line extending from $X$ to the first
interaction vertex $X_1$ (Fig.~2). Therefore, the $z\searrow 0$
behaviour of each diagram is dictated by the same function (apart from
potential IR problems), so that the analysis of the limit can be
essentially done in the first order. Nontrivial renormalization, in
contrast, becomes relevant only at higher order.  

\medskip

\hskip 35mm \begin{minipage}{120mm}
\epsfig{file=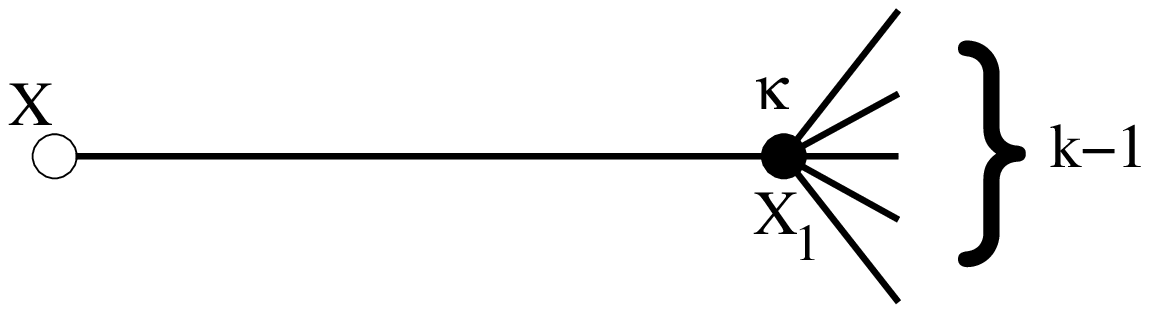,width=5.0cm}

{\bf Fig.~2:} Factorization of $\phi_{\kappa\LL}^{(n)}$. 
\end{minipage}

\medskip

To first order perturbation theory $n=1$ we obtain
\bea
\phi_{\kappa\phi^k}^{(1)}(X)=k \int _0^\infty \frac{dz_1}{z_1^{d+1}} \,
\gamma (z_1)
\int d^dx_1\, g(x_1)\cdot i\Delta_\ads^{\rm ret}(X,X_1)\;\wick{\phi^{k-1}(X_1)},
\label{Phi_L^1}
\eea 
where $\Delta^{\rm ret}_\ads(X,X_1) = (\Delta^+_\ads(X,X_1) -
\Delta^+_\ads(X_1,X))\theta(x^0-x_1^0)$ is the retarded propagator on 
$d+1$-dimensional AdS, according to \eqref{phi:kg} given by
\bea
\Delta^{\rm ret}_\ads(X,X_1) = \frac12(zz_1)^{d/2}
\int dm^2 J_\nu(mz)J_\nu(mz_1)\,\Delta^{{\rm ret}(d)}_m(x-x_1).
\label{Delta^ret_AdS}
\eea 
At this point, one might be tempted to read off
the $z\searrow0$ behaviour directly from \eqref{Delta^ret_AdS} and
the well-known behaviour of the Bessel functions near zero. We shall
see, however, that this attempt is too naive, and that the subsequent
$z_1$-integration in \eqref{Phi_L^1} changes the limit behaviour
substantially.   

\medskip

\subsection{Interaction $\LL=\kappa\phi$ (field shift)}
\label{sec:fieldshift}
For the trivial case $k=1$ (in which the ``interaction'' amounts
just to a shift of the field by a constant), the adiabatic limit
$\gamma(z_1)=1$, $g(x_1)=1$ can be taken directly in \eqref{Phi_L^1}
and yields the expected result   
\bea
\phi^{(1)}_{\kappa\phi}(X) = \int \frac{dz_1}{{z_1}^{d+1}}\int
d^dx_1\cdot i\Delta^{\rm ret}_\ads(X,X_1) = \frac 1{M^2} \ , 
\label{Phi^1:k=1}
\eea
which follows from $(\square_X+M^2)i\Delta^{\rm ret}_\ads(X,X_1)=
z^{d+1}\delta(z-z_1)\delta^d(x-x_1)$ upon integration over $X_1$, using
AdS-invariance so that the integral does not depend on $X$. One may
also perform the integrations explicitly in the representation
\eqref{Delta^ret_AdS} where the $x_1$-integration is obvious from
\eqref{D_m^ret-d}, and the subsequent $z_1$- and $m$-integrations are
carried out using formula (13.24(1)) in \cite{Wa},
\bea
  \int_0^\infty du\,{u^{\mu}}{J_\nu (u)} = 2^\mu\;
\frac{\Gamma(\frac12(1+\nu+\mu))}{\Gamma(\frac12(1+\nu-\mu))}\qquad
(-\nu-1 <\mu <\frac12) \ .
\eea
Clearly, the shift by a multiple of the ``constant field'' $1$
destroys the existence of the boundary limit with $z^{-\Delta}$. After
the subtraction of the vacuum expectation value (i.e., undoing the
shift), the boundary limit can be taken and reproduces the original
boundary field. This trivial example shows that in general,
interacting fields of different scaling dimensions may ``mix'', and
the appropriate boundary limits have to be taken after their separation.

\subsection{Interaction $\LL=\kappa\phi^2$ (mass shift)}
\label{sec:massshift}
In the case $k=2$, the interaction just amounts to a change of the AdS
mass by $\delta M^2=-2\kappa$, so that the perturbed field is just a
free field with a different mass. This is an instance of the
``Principle of Perturbative Agreement'' \cite{HW2}. Consequently, we
expect an anomalous dimension according to
$\Delta^\varphi_{\kappa\phi^2} = d/2 + \sqrt{(d/2)^2+M^2-2\kappa} =
\Delta-\kappa/\nu + \OO(\kappa^2)$ to arise. 
Thus, we are led to study the boundary
limit of
\bea
  \frac{\phi_{\kappa\phi^2}(z,x)}{z^{\Delta_{\kappa\phi^2}}}=
\frac{\phi(z,x)}{z^{\Delta}}+\kappa
\Bigl(\frac{\phi^{(1)}_{\kappa\phi^2}(z,x)}{z^{\Delta}}
+\frac1\nu\cdot\frac{\phi(z,x)}{z^{\Delta}}\>\log\>z \Bigr) 
+{\cal O}(\kappa^2)\;,
\label{log:anodim}
\eea
where the first order term \eqref{Phi_L^1} is
\bea
\label{Phi_1:k=2}
\phi^{(1)}_{\kappa\phi^2}(X) = 2\int \frac{dz_1}{{z_1}^{d+1}}\,\gamma(z_1)\int
d^dx_1\,g(x_1)\cdot i\Delta^{\rm ret}_\ads(X,X_1)\phi(X_1)\;.
\eea 
Indeed, in the partial adiabatic limit $\phi^{(1)}_{\kappa\phi^2}$
exhibits a logarithmic $z$-dependence which is precisely cancelled by
the combination occurring in \eqref{log:anodim}. Namely, 
\eqref{Phi_1:k=2} implies
\bea
(\square_X+M^2) \phi^{(1)}_{\kappa\phi^2}(X) = 2\gamma(z)g(x)\cdot\phi(X) 
\eea 
and consequently, using \eqref{laplace}
\bea
(\square_X+M^2) \Bigl(\phi^{(1)}_{\kappa\phi^2}(X)+\frac{\log z}\nu 
\cdot\phi(X)\Bigr) = \frac2\nu(\Delta-z\partial_z)\phi(X)
\label{inhomogen}
\eea 
in the region where $\gamma(z)=1$, $g(x)=1$. The right-hand side
vanishes in the limit $z\searrow0$ faster than $z^\Delta$ because the
leading $z^\Delta$ behaviour of the unperturbed field is annihilated
by the differential operator $\Delta-z\partial_z$. Since the
Klein-Gordon operator preserves homogeneity in $z$ (except for the
$z^2\square_x$ term which is suppressed at small $z$), the
combination of fields on the left-hand side also vanishes faster than
$z^\Delta$, up to a solution of the homogeneous equation. The
homogeneous solution can behave $\sim z^\Delta$ or $\sim z^{d-\Delta}$. 

If we can exclude the latter (dominant) contribution, then it follows
that the limit \eqref{log:anodim} at first order in $\kappa$ exists.  
Unfortunately, the previous argument based on the Klein-Gordon
operator cannot discriminate between $\sim z^\Delta$ and $\sim
z^{d-\Delta}$. We shall therefore develop a more refined analytical
method of computation which is ``universal''  (see Lemma~B.1 in
App.~\ref{app:lemma}) in the sense that it can also be applied when
dealing with interactions of higher polynomial degree
(Sect.~\ref{sec:polynomial}) and with diagrams with loops
(Sect.~\ref{sec:phi2}). This method at the same time shows the
emergence of the $z^\Delta \,\log z$ terms. The argument is lengthy,
with essential parts contained in App.~\ref{app:lemma}, but it is
crucial for the understanding of the boundary limit. 

For the sake of transparency and  computational simplicity, we present
only the case  
\bea
d=3 \qquad\hbox{and}\qquad M=0 \ .
\eea
The AdS 2-point functions are explicitly known in terms of
hypergeometric functions or associated Legendre functions of the
second kind \cite{F,BBMS}: Let $X=(z,x)$, $X_1=(z_1,x+y)$
($z,z_1\in\RR_+$; $x,y\in\MM^3$), and  
\bea
v=\frac{z^2+z_1^2-y^2}{2zz_1}\ .
\eea 
$v$ is AdS-invariant. Namely, viewing $\ads_{d+1}$ as the
hypersurface $\xi\cdot\xi=1$ in a $d+2$-dimensional ambient space
$\RR^{d+2}$ of signature $(+,-\ldots-,+)$, we have 
\bea
v=\xi\cdot \xi_1\ ,
\eea 
hence $v$ is related to the ``chordal distance'' by
$d(\xi,\xi_1)=(\xi-\xi_1)^2=2(1-v)$. We expect singularities at
$d(\xi,\xi_1)=0$ ($\Leftrightarrow v=1$) and, due to the the
identification of $-\xi_1$ with $\xi_1$, also at $d(\xi,-\xi_1)=0$
($\Leftrightarrow v=-1$). Note also that timelike separation between
$X$ and $X_1$ corresponds to $v\in [-1,1]$. Then for $d=3$  
\bea
\label{Delta_ads}
\Delta^+_\ads(X_1,X) = -\frac{1}{4\pi^2} \; Q'_{\nu-\frac
  12}(v+iy^00)\ .
\eea 
Here $Q_\ell(u)$ is a solution of Legendre's differential equation 
\bea
(1-u^2)f'' - 2u f' + \ell(\ell+1)f=0 \ ,
\eea
which is analytic outside a cut along the real interval $[-1,1]$. For $M=0$,
hence $\nu=\frac32$, $\Delta=3$, it is the elementary function
\bea
\label{Q1}
Q_1(u)= \frac u2\log\frac{u+1}{u-1}\;-1 \qquad\Rightarrow\qquad Q_1'(u)=\frac
12(1+u\partial_u)\log\frac{u+1}{u-1}.
\eea 
The retarded propagator $\Delta^{\rm ret}_\ads(X,X_1)
=\big(\Delta^+_\ads(X,X_1)-\Delta^+_\ads(X_1,X)\big)\theta(-y^0) $ is
given by the discontinuity across the cut:
\bea
4\pi i\,\Delta^{\rm ret}_\ads(X,X_1)&=& 
\frac
1{2\pi}(1+u\partial_u)\log\frac{u+1}{u-1}\Big\vert^{u=v+i0}_{u=v-i0}\cdot
\theta(-y^0) \notag \\ &=& -(1+v\partial_v)
\theta(1-\vert v\vert)\cdot \theta(-y^0).
\eea
This discontinuity is to be understood as a distribution by partial
integration w.r.t.\ $v$:
\bea
\label{hdef}
H[f]:= -\int dv \,f(v)(1+v\partial_v)\theta(1-\vert v\vert) =
\int_{-1}^{+1} dv\, v\partial_v f(v).
\eea

Because we have represented the retarded propagator as a
distribution w.r.t.\ the variable $v$, we have to perform all other
integrations (at fixed value of $v$) first. We therefore change the
integration variables: in spatial polar coordinates, let 
$y = (-t,r \,\vec e_\varphi)$, and $w:= y^2 \equiv t^2-r^2$. Then the
new variables are 
\bea
v\equiv \frac{z^2+z_1^2-w}{2zz_1},\quad z_1,\quad t \equiv
-y^0,\quad \varphi. \eea
The measure becomes
\bea
d^3y\,\theta(-y^0) \,\frac{dz_1}{z_1^4}\,\theta(z_1) = z\cdot
dv\cdot \frac{dz_1}{z_1^3} \, \theta(z_1) \cdot
dt\,\theta(t)\,\theta(t^2-w)\cdot d\varphi,
\eea
where 
\bea
\label{wz}
w = w_{v,z}(z_1)= z^2+z_1^2-2v\cdot zz_1 
\equiv (z_1v-z)^2 + (1-v^2)z_1^2.
\eea

There is a dense domain of vectors for which matrix elements
$(\Psi_1,\phi(X_1)\Psi_2)$ of the distributional field become
smooth function. We then extract the leading $z_1$ behaviour  
and write 
\bea
\Gamma(z_1,x^0-t,\vec x+r\vec
e_\varphi):=\gamma(z_1)g(x_1)\cdot z_1^{-3}\,
(\Psi_1,\phi(z_1,x_1)\Psi_2). 
\eea
This is a smooth function with compact support, because of the cutoff
functions $g$ and $\gamma$. At $z_1=t=r=0$, it equals the corresponding
matrix element of $\varphi(x)$, because $g(x_1)=1$ and $\gamma(z_1)=1$
in the region of interest (partial adiabatic limit).  
Finally we average over the spatial directions and put
\bea
\Gamma_x(z_1,t,r^2):= \frac 1{2\pi}\oint d\varphi\; 
\Gamma(z_1,x^0-t,\vec x+r\vec e_\varphi). 
\eea
Then $\Gamma_x$ is smooth
\footnote{It will be important later (App.~\ref{app:lemma}) that
  $\Gamma_x$ is regular in the {\em quadratic} variable $r^2$. This is
  obvious at $r>0$ because the square root is smooth. At $r=0$, the
  smoothness can be seen by a Taylor expansion with remainder of
  $\Gamma(z_1,x^0-t,\vec x+r\vec e_\varphi)$, because the angular
  averaging annihilates all odd terms.}  
in all three arguments $\geq 0$, and 
\bea
\label{gx0}
\Gamma_x(0,0,0) = (\Psi_1,\varphi(x)\Psi_2). 
\eea

With these preparations, (the matrix element of) the first-order
correction \eqref{Phi_1:k=2} to the renormalized field becomes   
\bea
\label{HGamma}
(\Psi_1,\phi^{(1)}_{\kappa\phi^2}(X)\Psi_2)  
= z\cdot H\Big[\int_0^\infty \!\!dz_1 \int_0^\infty \!\!dt\,
\theta(t^2-w)\cdot 
\Gamma_x(z_1,t,t^2-w)\big\vert_{w=w_{v,z}(z_1)}\Big]\;, \quad
\eea
with the functional $H[\cdot]$ as defined in \eqref{hdef}. We claim,
that this equals   
\bea 
(\Psi_1,\phi^{(1)}_{\kappa\phi^2}(X),\Psi_2)
 = -\frac 23\, z^3\Big(\log z\cdot \Gamma_x(0,0,0) + \hbox{(regular) }\Big),
\label{claim}
\eea 
where (regular) stands for a contribution that is regular in $z$ at $z=0$.

The argument goes as follows. For a smooth function $f$ on $\RR^3$
with compact support, we denote by $I_0(v,z)(f)$ the integral
\bea
\label{i0vz} I_0(v,z)(f):= \int_0^\infty dz_1 \; \int_0^\infty
dt\,\theta(t^2-w)\; f(z_1,t,t^2-w)\big\vert_{w=w_{v,z}(z_1)} .
\eea
Thus, to compute \eqref{HGamma}, we have to apply the functional $H$ to
$I_0(v,z)(f)$ when $f$ equals $\Gamma_x$ on $\RR_+^3$. 

In App.~\ref{app:lemma}, we prove that $I_0(v,z)(f)$ is continuous
w.r.t.\ $v$ and differentiable in the range $v^2<1$. Thus, the
definition \eqref{hdef} of $H$ by partial integration is unambiguous,
and it is sufficient to know this function at $v^2<1$, where
$w\geq (1-v^2)z^2 >0$. In physical terms, this remark means that there
are no singular contributions from lightlike $y$ ($w=0$): the integration
\eqref{Phi_1:k=2} can be properly computed by exhausting the backward
lightcone ``from the inside''.  

In App.~\ref{app:lemma}, we also prove that in the range $v^2<1$,
$I_0(v,z)(f)$ is of the form 
\bea
I_0(v,z)(f)= \hskip-2mm\sum_{0\leq k\leq\ell \leq 2} \hskip-3mm
A_{k\ell}(f)\,v^k\,z^\ell + z^2\cdot\frac {1-v^2}2\,
\log \big((1-v)z\big)\cdot f(0,0,0) + R_{v,z}(f)\quad 
\eea
where $A_{k\ell}$ are certain distributions that do not depend on $v$ and
$z$, while the remainder $R_{v,z}$ is a family of distributions that
is differentiable w.r.t.\ $v$ in the range $v^2<1$, and vanishes 
$\sim z^3$ at $z=0$. 

Noting that $H[v^0]=H[v^1]=0$, the leading terms are annihilated: 
\bea
H\big[I_0(v,z)(f)\big]& =& -\frac{H[v^2]}2\,z^2\,\log z \cdot
f(0,0,0)+ \\ &+&
z^2\Big(A_{22}(f) + H\big[\frac{1-v^2}2\log(1-v)\big]\cdot f(0,0,0)\Big) + H[R_{v,z}(f)]. \notag 
\eea
Thus, with $H[v^2]=\frac 43$, we have

\medskip

\noindent {\bf Proposition 3.1:} {\sl For any test function $f$ on $\RR^3$,
  the limit
\bea
\lim_{z\searrow0} z^{-2} \Big\{H
\big[I_0(v,z)(f)\big] +
\frac 23\,z^2\log z \cdot f(0,0,0) \Big\} \quad
\eea
is finite.} 

\medskip

For $f=\Gamma_x$ on $\RR_+^3$, this is our claim \eqref{claim}. This
ensures that $\phi^{(1)}_{\kappa\phi^2}(X)$ decays at least like $z^3\log z$, 
and because of \eqref{gx0}, it also ensures that
$\phi^{(1)}_{\kappa\phi^2}(X)+\frac{\log z}\nu \cdot\phi(X)$ (recall
$\nu=\frac32$ in \eqref{log:anodim}) decays at least like $z^\Delta=z^3$.  
In other words, the boundary limit exists (in first order perturbation
theory, and in the obvious weak sense),
and is exactly given by the expected correction of the scaling
dimension of the boundary field.   

\medskip

Apart from establishing the existence of the (expected) boundary
limit, the main message to be drawn from the computation in
App.~\ref{app:lemma}, however, is that 
\begin{itemize}
\item\it the origin of the logarithmic term (corresponding to the
  anomalous dimension) is the range $z_1=0$ of the integral
  \eqref{Phi_1:k=2}, and not the power law behaviour of the retarded
  propagator at $z=0$. 
\end{itemize}

\subsection{Interactions $\LL=\phi^k$ ($k>2$)}
\label{sec:polynomial}
We now turn to the non-trivial interactions $k>2$. 
In these cases \eqref{Phi_L^1} yields 
\bea
(\square_X+M^2) \phi^{(1)}_{\kappa\phi^k}(X) = \gamma(z)g(x)\cdot k
\wick{\phi^{k-1}(X)} 
\eea 
where the right-hand side $\sim z^{(k-1)\Delta}$ vanishes faster than
$z^\Delta$. By the same argument as used after \eqref{inhomogen},
$\phi^{(1)}_{\kappa\phi^k}(X)$ behaves either like $z^\Delta$ or like
$z^{d-\Delta}$. In the special case $d=3$, $M=0$, we can explicitly
see the absence of the ``wrong'' contribution $\sim z^{d-\Delta}$, by
repeating the explicit computation as in the previous section. 
Replacing $\phi(X_1)$ by $\wick{\phi(X_1)^{k-1}}$, one gets an
additional factor $z_1^{3(k-2)}$ in \eqref{i0vz}. Because the 
logarithmic term in this case appears at order $\OO(z^{3+3(k-2)})$ 
(Lemma~B.1), it is manifest that the first-order term is of order
$\OO(z^3)$, as desired, and there is no logarithmic term. Thus, the
boundary limit exists without an anomalous dimension.  

Although a complete analysis of renormalization at higher-order is
beyond the scope of this paper, let us anticipate what happens in the
case at hand. First, we observe (see Fig.~2
above) that $\phi_{\kappa\phi^k}^{(n)}$ can be written as 
\bea
\phi_{\kappa\phi^k}^{(n)} = k \int_0^\infty \frac{dz_1}{z_1^{d+1}}
\gamma (z_1)
\int d^dx_1\, g(x_1)\cdot i \Delta^{\rm ret}_\ads(X,X_1) \,
\big(\phi^{k-1}\big)_{\kappa\phi^k}^{(n-1)}(X_1). 
\eea  
Thus, in order to renormalize $\phi_{\kappa\phi^k}$ at order $n$, one 
previously has to renormalize $\big(\phi^{k-1}\big)_{\kappa\phi^k}$ at
order $n-1$. In principle, one has to renormalize ``all fields
simultaneously'', but in practice, for any finite order of any given
field it is sufficient to renormalize only a finite number of fields
to lower orders.  

Thus, assuming recursively that $\left(\phi^{k-1}\right)_{\kappa\phi^k}$ 
has been defined up to $(n-1)$st order, and anticipating that its
boundary limit exists with an anomalous dimension of order $\OO(\kappa)$,
then $\left(\phi^{k-1}\right)_{\kappa\phi^k}^{(n-1)}$ behaves like
$z^{(k-1)\Delta}$ times a polynomial in $\log z$, as $z\searrow 0$. 
Because the canonical dimension $(k-1)\Delta$ is larger than $\Delta$, 
the same argument as before applies to ensure that the partial
adiabatic limit for $\phi_{\kappa\phi^k}$ is unproblematic, and for
$z$ sufficiently small (such that $\gamma(z)=1$), the equation 
\bea
(\square_X+M^2) \phi^{(n)}_{\kappa\phi^k}(X) = k\cdot g(x)\, 
\big(\phi^{k-1}\big)_{\kappa\phi^k}^{(n-1)}(X)
\eea 
implies the $z^\Delta$ behaviour of $\phi^{(n)}_{\kappa\phi^k}(X)$ as
$z\searrow 0$. Again, this equation does not yet exclude a
term $\sim z^{d-\Delta}$, but an explicit computation as in Lemma~B.1
in the special case $d=3$, $M=0$ again shows its absence. We conclude   
that anomalous dimensions do not arise also in higher orders of
perturbation theory.   

Actually, one can go beyond this statement: even if the logarithms
could be summed (borrowing suitable higher order terms, i.e.,
violating the proper perturbative systematics) to give rise
to an anomalous dimension $\Delta^{\phi^{k-1}}_{\kappa\phi^k}$ up to
order $n-1$ (see Sect.~\ref{sec:phi2}), then the argument would still
hold true as long as $\Delta^{\phi^{k-1}}_{\kappa\phi^k} > \Delta$
(cf.\ Lemma~B.1 with $n=\Delta^{\phi^{k-1}}_{\kappa\phi^k} - \Delta$). 

In the next section, we shall discuss the behaviour of ``composite
fields'' $\big(\phi^2\big)_{\kappa\LL}$. Depending on the interaction,
these fields will exhibit finite anomalous dimensions.

\subsection{Comparison of bulk vs boundary renormalization schemes}
\label{sec:comparison}
We conclude this section with a comparison of the competing
renormalization prescriptions in the case at hand. Concerning the
renormalization, we find here significant differences between (a) our
procedure, as just outlined, and (b) perturbation theory around the
generalized free field $\varphi$ in Minkowski space $\MM_d$, requiring
Poincar\'e invariance (b1), or in conformal Minkowski space $\CM_d$,
requiring conformal invariance (b2):  
\begin{itemize}
\item[(a)] (Renormalization in the bulk) The numerical distribution 
  $r^\circ(X_1;X) = (\Omega,R_{1,1}($ $\phi(X_1);\phi(X))\Omega)$
  coincides with the retarded propagator $i\Delta^{\rm ret}_\ads(X,X_1)$ 
  in $\ads$. Its extension to the diagonal is uniquely given by
  \eqref{Delta^ret_AdS}, and there is no freedom of renormalization, 
  because its scaling degree in the relative coordinates equals 
  $d-1$ (for $z>0$), which is smaller than the dimension of the
  relative coordinates ($=d+1$) \cite{BF}. The boundary behaviour of
  the resulting fields is dominated by the $z_1$-integration near
  $z_1=0$, which depends sensitively on the operator valued
  distribution with which $r$ is multiplied. It is important to keep
  in mind that we have {\it renormalized (extended $r^\circ$ to the
    diagonal) first, and then taken the limit $z\searrow 0$ (in the
    partial adiabatic limit at the boundary)}.   
\item[(b)] (Renormalization on the boundary) Doing perturbation theory
  on the boundary, instead, we have to take the limit $z\searrow 0$
  first. This yields the unrenormalized distribution
  $r_\varphi^\circ(x-x_1) = (\Omega,R_{1,1}(\varphi(x_1);\varphi(x))\Omega)$:
\bea
r_{\varphi}^\circ(x-x_1) = i[\varphi(x),\varphi(x_1)]\theta(x^0-x_1^0)
=\int dm^2 \, m^{2\nu}\, i\Delta_m^\mathrm{ret}(x-x_1).
\label{retprop:M}
\eea 
This product of distributions exists on ${\cal D}(\MM_d)$ only in the
range $-1<\nu <0$ 
\footnote{The expression on the 
  right side results from the definition of $\varphi$
  \eqref{gff}. Alternatively, it can be obtained by taking the
  boundary limit $\lim_{z,z_1\searrow 0}(zz_1)^{-\Delta}\ldots$
  \eqref{boundlim} of \eqref{Delta^ret_AdS}. This limit may be done
  before the mass integration in \eqref{Delta^ret_AdS} iff $-1<\nu <0$.}.
For $\nu\geq 0$ the integral $\int dm^2\frac{m^{2\nu}}{m^2-p^2-ip^00}$
diverges, nevertheless $[\varphi(x), \varphi(x_1)]\theta(x^0-x_1^0)$
is well defined for $x\not= x_1$, and one is faced with the problem to
extend $r_{\varphi}^\circ$ from ${\cal D}(\MM_d\setminus\{0\})$ to
${\cal D}(\MM_d)$. One has two options:
\begin{itemize}
\item Case (b1): One only requires that the Lorentz invariant
  extension does not increase the scaling degree (with respect to $0$)
  of $r_{\varphi}^\circ$ \cite{EG,BF}, which has the value 
  ${\rm sd}(r_{\varphi}^\circ)=2\Delta =d+2\nu$. In this case, the
  retarded propagator is {\it non-unique} for $\nu\geq 0$: 
  the general solution reads
\bea
     r_{\varphi}(y)=(\mu^2-\square_y)^{[\nu]+1}\int dm^2\frac{m^{2\nu}
i\Delta_m^\mathrm{ret}(y)}{(\mu^2+m^2)^{[\nu]+1}}+\sum_{n\leq \nu}
C_n\square^n\delta(y)
\label{retprop:M-int}
\eea
  where $\mu >0$ and the $C_n$'s are arbitrary constants (cf.\ Appendix C
  of \cite{DF3}). Clearly, the renormalization mass $\mu$ and the local
  terms break the scale invariance (unless $n=\nu$). 
\item Case (b2): Requiring conformal covariance of the extension, a
  necessary condition is that the homogeneous scaling behaviour of
  $r^{\varphi}$ is maintained: this is an intensification of the
  requirement in (b1). From \eqref{retprop:M-int} we see that there is
  a unique solution for $-1<\nu\not\in\NN_0$ which is obtained by
  choosing $\mu=0$ and $C_n=0\>\forall n$. But if $\nu\in\NN_0$, the
  mass integral is IR-divergent for $\mu=0$, and a scaling covariant
  retarded propagator does not exist.  
\end{itemize}
\end{itemize}

\section{Case studies II: The interacting composite field
  $(\varphi^2)_{\kappa\phi^k}$ }  
\label{sec:phi2}
\setcounter{equation}{0} 
\subsection{General considerations}
\label{sec:general}
We turn to the field $(\phi^2)_{\kappa\LL}$ with interaction
$\LL=\wick{\phi^k}$ ($k\geq 2$). In this case, there exist three types of
diagrams which a priori behave differently as $z\searrow 0$: those
diagrams in which the two interaction vertices connected to the field
vertex are distinct and do not belong to a common loop, those in which
they are distinct and belong to a common loop, and those in which they
coincide.  

\vskip5mm
\indent\begin{minipage}{120mm}
\hskip3mm\epsfig{file=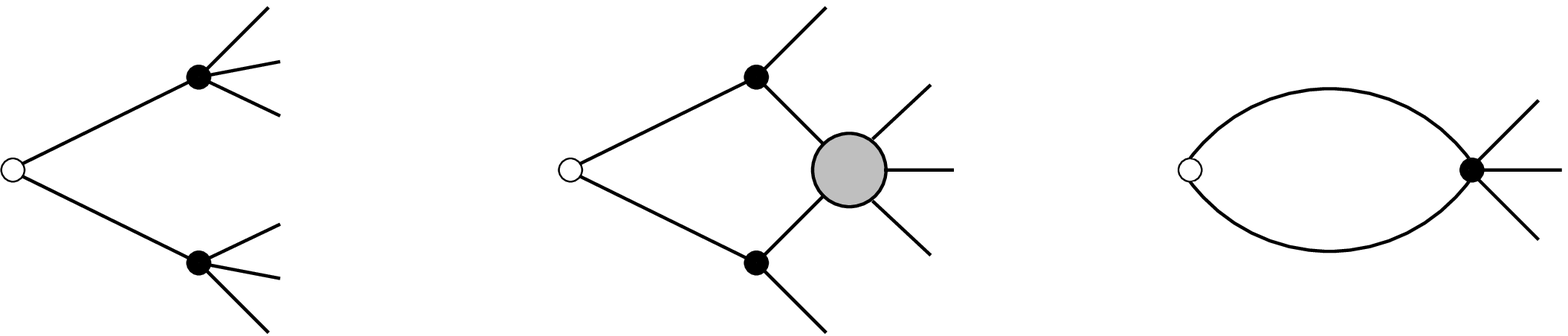,width=11.5cm}

\small{\bf Fig.~3: Three types of diagrams} arising in 
perturbation theory for the interacting field
$\big(\phi^2\big)_{\kappa\LL}(X)$ with interaction $\LL=\wick{\phi^k}$.
\end{minipage}
\vskip5mm

Diagrams of the first type factorize into two diagrams as for the
field $\phi_{\kappa\LL}$ and consequently can be treated as in
Sect.~\ref{sec:polynomial}. The second type does not arise in first
order. Diagrams of the last type contain the fish diagram (Fig.~1) as
a subdiagram, which determines their $z$-dependence. This diagram
gives the contribution to 
$(\phi^2)_{\kappa\phi^k}$ 
\bea
\frac{k(k-1)}{2}\int_0^\infty
\frac{dz_1}{z_1^{d+1}}\;\gamma (z_1)\int dx_1\, g(x_1)\;
r_\fish(X_1;X)\wick{\phi^{k-2}(X_1)} \ .
\label{fish:intf}
\eea
In order to define this contribution, the unrenormalized distribution
$r^\circ_\fish(X_1;X)\equiv
(\Omega,R^\circ_{1,1}(\phi^2(X_1);\phi^2(X))\Omega)$, given by  
\bea
  r_\fish^\circ(X_1;X)\equiv -2i\left(\Delta^{+}_\ads(X_1,X)^2 -
    \Delta^{+}_\ads(X,X_1)^2\right)\theta(x^0-x_1^0)
\label{fish:nonren}
\eea
at $X_1\neq X$ (cf.\ \eqref{typical}), has to be extended to the
diagonal $X_1=X$. Then we have to study the boundary behaviour $z\searrow0$
of the renormalized integral \eqref{fish:intf} in the partial
adiabatic limit. Our task is to understand the influence of the UV
renormalization on the boundary limit.   

The unrenormalized distribution \eqref{fish:nonren}
is real-valued and AdS-invariant. We require
that the extension $r_\fish(X_1;X)$ has the same properties:
\begin{itemize}
\item[(I)] $r_\fish$ is real-valued (i.e., $r_\fish
  (f)^*=r_\fish(f^*)$) and the scaling degree in the
  relative coordinates $Y=(y,u)$ is not increased by the extension:
\bea
    \mathrm{sd}_{Y}(r_\fish(\>\cdot\> ;X))=\mathrm{sd}_{Y}
  (r_\fish^\circ(\>\cdot\> ;X))=2d-2\quad\forall X.
\label{fish:sd}
\eea
\item[(II)] $r_\fish$ is AdS-invariant
\bea
   r_\fish(tX_1;tX)=r_\fish(X_1;X)\quad\forall t\in  SO(2,d)\ .
\label{ads-inv}
\eea
\end{itemize}
In addition, we want to impose the existence of the boundary limit of
the interacting field $(\phi^2)_{\kappa\phi^k} = \wick{\phi^2} +
\kappa (\phi^2)_{\kappa\phi^k}^{(1)} + \OO(\kappa^2)$ as a
condition on the renormalization, admitting for an anomalous
dimension $2\Delta + \kappa\delta +
\OO(\kappa^2)$. 

Thus, up to first order of perturbation theory, 
\bea
\frac{\wick{\phi^2(z,x)}}{z^{2\Delta}}
+\kappa\Bigl(\frac{(\phi^2)^{(1)}_{\kappa\phi^k}(z,x)}{z^{2\Delta}} 
-\delta\frac{\wick{\phi^2(z,x)}}{z^{2\Delta}}\>\log\>z\>\Bigr)
\label{log:cancel}
\eea
should converge with $z\searrow0$. We have already seen that the
contributions from the first type of diagrams (Fig.~3) to
$(\phi^2)_{\kappa\phi^k}^{(1)}$ behave $\sim z^{2\Delta}$ if $k>2$,
and with a logarithmic correction if $k=2$, so that their limit exists
separately because $\Delta>0$. Because the only possibly divergent
contribution comes from the fish diagram integrated with
$\wick{\phi^{k-2}}$, a cancellation against the contribution from an
anomalous dimension can occur in \eqref{log:cancel} only if $k=4$ and
only if the divergence of $z^{-2\Delta}r_\fish(X_1,X)$
integrated with $\wick{\phi^{2}}$ is logarithmic. Thus, we are led to
require    
\begin{itemize}
\item[(III)] The renormalized expression \eqref{fish:intf} taken in
  the partial adiabatic limit and multiplied by $z^{-2\Delta}$
  converges at $z\searrow0$ if $k\neq 4$, while for $k=4$ it may diverge
  $\sim \log z\frac{\wick{\phi^2(X)}}{z^{2\Delta}}$. 
\end{itemize}

Due to general theorems \cite{BF,HW} there exist extensions which
fulfill (I) and (II). For $d\leq 4$, these two requirements reduce the
freedom of normalization to
\bea
  r_\fish(X_1;X)+Cz^{d+1}\delta (x_1-x)\delta(z_1-z)\ .
\label{freiheit}
\eea
So there is only one normalization constant $C$ at disposal to fulfil (III). 
For this reason, we concentrate on $d= 3$ and $d=4$ from now on. 

Changing the value of $C$ just adds a multiple of $\wick{\phi^{k-2}}$
to $(\phi^2)_{\kappa\phi^k}^{(1)}$. If $k=2$ or $k=3$, this term
$\sim z^0$ or $\sim z^\Delta$ must not be present in the boundary limit 
taken with $z^{-2\Delta}$, so condition (III) -- if it can be fulfilled -- 
fixes the value of $C$, and thus determines a ``field mixing''. If
$k=4$, the addition just amounts to a multiplicative renormalization
of the zero order term. If $k>4$, the addition is ineffective in the
boundary limit. In both cases $k\geq 4$, the renormalization parameter
$C$ is unconstrained by condition (III). These a priori conclusions
are in perfect agreement with the corresponding conclusions drawn from
the analysis of Witten diagrams for correlation functions in the dual
approach to the AdS-CFT correspondence \cite{Witten}.

\subsection{$d=3$, $M=0$: Renormalization of the fish diagram on
  $\ads_4$} 
\label{sec:3fish4}
The standard strategy \cite{HW,HW2} to renormalize (extend) a distribution
like the fish diagram $r_\fish^\circ$ in curved space-time is to
pass to the scaling limit $\bar r_\fish^\circ$ which gives a
distribution in the tangent space at the point $X$. The latter carries
the leading UV singularity and can be renormalized as in flat space
(with the constant metric $g_X$), while the less singular ``reduced'' 
distribution $r_\fish^{\circ \rm red} = r_\fish^\circ - \bar
r_\fish^\circ$  is (in $d=3$ or $d=4$) uniquely extended ``by
continuity''. The problem with this strategy in our situation is that
$r_\fish^{\rm red}$ and $\bar r_\fish$ (the latter being
independent of $\Delta$ because the scaling limit looses the
information about the AdS mass $M^2$) behave differently at the
boundary, and do not allow us to deduce the boundary behaviour of the
integral \eqref{fish:intf}. 

Let us look more closely at the distribution \eqref{fish:nonren}.
Unfortunately, the AdS K\"allen-Lehmann expansion of $(\Delta^+_\ads)^2$
is not known explicitly \cite{BEM}, with which one could perform the
renormalization in the spirit of \eqref{flatrenorm}. Instead, we shall
use again the explicit form \eqref{Delta_ads} of
$\Delta^+_\ads(X_1,X)\propto Q_{\nu-\frac12}'$ in $d+1=4$ bulk
dimensions, and its elementary expression \eqref{Q1} if $M=0$, hence
$\nu=\frac32$ and $\Delta=3$. 

In order to renormalize \eqref{fish:nonren} (i.e., to define the
retarded product as a distribution on $\ads_4^{\times 2}$), we adopt
the method of differential renormalization \cite{FJL}: As a
distribution on $\ads_4^{\times 2}\setminus\{(X,X)\vert X\in\ads_4\}$,
\eqref{fish:nonren} is of the form
\bea
r^\circ_\fish(X_1,X) = j(X_1,X)\theta(x^0-x_1^0) 
\eea
with $j(X_1,X)\propto
Q_{\nu-\frac12}'(v-iy^00)^2-Q_{\nu-\frac12}'(v+iy^00)^2$. One writes  
\bea
j(X_1,X)=\square_{X_1}J(X_1,X)
\eea
where $J$ is an $\ads$-invariant distribution which vanishes if $X_1$
is spacelike separated from $X$, and $\mathrm{sd}(J) < \mathrm{sd}(j)$, 
so that $J(X_1,X)\theta(x^0-x_1^0)$ is well-defined as a distribution on
$\ads_4^{\times2}$. One then defines    
\bea
r_\fish(X_1,X) :=
\square_{X_1}\big(J(X_1,X)\theta(x^0-x_1^0)\big).
\eea
At $X\neq X_1$, this differs from the unrenormalized distribution 
$\theta(x^0-x_1^0)\cdot\square_{X_1}J(X_1,X)$ by a term
\bea
\propto \partial_0\big(J(X_1,X)\delta(x^0-x_1^0)\big) +
\delta(x^0-x_1^0) \partial_0 J(X_1,X).
\eea
The support property of $J$ ensures that this vanishes at $X\neq X_1$,
hence $r_\fish(X_1,X)$ is indeed an extension of $r^\circ_\fish$. 
Obviously, $r_\fish$ satisfies the requirements (I) and (II) above.

We follow this strategy in the case $M=0$, where by \eqref{Delta_ads},
$\Delta^+_\ads(X_1,X)\sim Q_1'$ is given explicitly in terms of the
elementary function \eqref{Q1}. We thus obtain 
\bea
(\Delta^+_\ads(X_1,X))^2 = \hspace{105mm} \\
= \frac{1}{64\pi^4}\;\Big(\big(\log\frac{u+1}{u-1}\big)^2+ u\partial_u
\big(\log\frac{u+1}{u-1}\big)^2 + \big(\frac{u}{u+1}-
\frac{u}{u-1}\big)^2\Big)\big\vert_{u=v+iy^00}.
\notag \eea
Here, the first term is a logarithmically bounded function, hence
well-defined as a distribution, and consequently also the second. 
The last term is defined as a distribution by  
\bea
\big(\frac1{v\pm1+iy^00}\big)^2 = -\partial_v
\big(\frac1{v\pm1+iy^00}\big).
\eea

We now look for a {\em function} $F$ such that 
\bea
\label{c1}
\square_{X_1} F(u) \equiv (1-u^2)F''(u)-4u F'(u) = Q_1'(u)^2
\eea 
and
\bea
\label{c2}
J(X_1,X)=\frac
{i}{8\pi^4}\big(F(v-iy^00)-F(v+iy^00)\big)=0\qquad\hbox{if}\quad
\vert v\vert >1.
\eea 
Next, we determine the discontinuity along the cut 
\bea
\delta F(v) = F(v+i0) - F(v-i0)
\eea 
as a {\em distribution}. Then, we can define the renormalized fish diagram as
\bea
\label{fish:ren}
r_\fish(X_1,X)=\frac {i}{8\pi^4}\cdot
\square_{X_1}\big(\delta F(v)\theta(-y^0)\big) . 
\eea 

\medskip

\noindent {\bf Proposition 4.1:} {\sl Equation \eqref{c1} is solved by
\bea\label{F(u)}
F(u)&=&\frac 1{2}\Bigl(\Li_3\frac{2}{1-u}+\Li_3\frac{2}{1+u}\Bigr)+
\frac 1{6}\,\frac{d}{du}\Bigl(\Li_3\frac{2}{1-u}-\Li_3\frac{2}{1+u}\Bigr)
\notag\\
&&+\frac 16 \,\log\,\frac{u+1}{u-1}\,\cdot\,\Bigl(\Li_2\frac{2}{1+u}
-\Li_2\frac{2}{1-u}\Bigr)-\frac 1{16} \,\Bigl(\log\,\frac{u+1}{u-1}\Bigr)^2\\
&&+\frac 1{144}\,(u^2+3)\,\frac{d}{du}\Bigl(\log\,\frac{u+1}{u-1}\Bigr)^3
+\frac {u}{16}\,\frac{d}{du}\Bigl(\log\,\frac{u+1}{u-1}\Bigr)^2
+\frac{6-u^2}{12(u^2-1)}\qquad
\notag
\eea 
plus the general solution $C_1 Q_1'(u) + C_2$ of the
homogeneous equation.}

\newpage

We point out that $F(u)$ is analytic for $u\in\CC\setminus [-1,1]$
(see Appendix~\ref{app-pf}),  
and that the particular solution given by the Proposition
is symmetric ($F(-u)=F(u)$), but $Q'_1(-u)=-Q'_1(u)$. By writing some
terms as derivatives, the boundary values $F(v\pm i\varepsilon)$ are defined
as distributions. 

\medskip 

{\it Proof:} by insertion into \eqref{c1}. In Appendix~\ref{app-pf} we
sketch the derivation of \eqref{F(u)}.  


\medskip

\noindent {\bf Proposition 4.2:} 
{\sl The discontinuity $\delta F(v) = F(v+i0)-F(v-i0)$ is given by 
\bea\label{dF}
\delta F(v) = i\pi\Big(\theta(1-\vert v\vert)\,h_0(v) + \partial_v
\big(\theta(1-\vert v\vert)\,h_1(v)\big)\Big),
\eea
where }
\bea 
h_0(v) &=& \frac 13\Big(\Li_2\frac{1+v}2 - \Li_2\frac{1-v}2\Big)
+ \frac 23\log\frac{1+v}{1-v} \, ,
\notag \\
h_1(v)&=& \frac 13 \log\frac{1+v}2\log\frac{1-v}2 
-\frac v4\log\frac{1+v}{1-v} + \Big(\frac{\pi^2}{18}+\frac5{12}\Big).
\eea

\medskip

Notice that the derivative of $\theta(1-\vert v\vert)$ cannot be taken
separately, because $h_1$ is logarithmically divergent at
$v=\pm1$. Instead, $\delta F$ is understood as a distribution in $v$,
where the derivative is defined by partial integration, see below.

\medskip 

{\it Proof:} See Appendix~\ref{app-pf}.


\medskip

Adding the homogeneous solutions, the second of the integration
constants, $C_2$, does not contribute to the discontinuity. Thus, the
(expected) renormalization freedom consists in adding to
\eqref{fish:ren} the term   
\bea\label{renfree}
&&\frac{i}{8\pi^4}
\,C_1\,\square_{X_1}\,\Bigl(\bigl(Q'_1(v-iy^00)-Q'_1(v+iy^00)\bigr)\, 
\theta(-y^0)\Bigr) 
= \\ 
&&\qquad\qquad= \frac{-i}{2\pi^2}\,C_1\,\square_{X_1}\,\Delta^{\rm ret}_\ads(X,X_1) = -\frac{C_1}{2\pi^2}\,z^4\,\delta(z_1-z)\,\delta^{(3)}(x_1-x)\ .
\notag\eea
{\bf Remarks:} (i) In contrast to the renormalization of the massless
fish diagram in 4-dimensional Minkowski space, the present {\em
  renormalization on AdS does not require the introduction of a mass
  scale.} This is because there is already a mass scale in the
formalism, namely $1/R^2$, where $R$ is the radius of AdS. (In our
conventions: $R^2\equiv
(\xi^0)^2-\sum_{k=1,2,3}(\xi^k)^2+(\xi^4)^2=1$.)\\ 
(ii) $\delta F$ in Prop.~4.1 is {\em antisymmetric} in $v$; however the
renormalization freedom \eqref{renfree} is {\em symmetric} in
$v$. Hence, there is a {\em distinguished renormalization}: $C_1=0$. 

\medskip

The term \eqref{renfree} contributes a multiple of
$\wick{\phi^{k-2}(X)}$ to the first order term of $(\phi^2)_{\kappa\phi^4}$. 
As discussed in Sect.~\ref{sec:general}, for $k>4$ this terms
does not contribute to the boundary limit, while for $k=4$ its
boundary limit $\wick{\varphi^2(x)}$ exists trivially and amounts 
to a multiplicative renormalization of $(\varphi^2)_{\kappa\phi^4}$. 
For $k=3$, it produces a ``mixing'' of the field $\wick{\phi^2}$ with
$\phi(X)$, and the boundary limit has to be taken of the appropriate
mixed field (cf.\ the end of the Sect.~\ref{sec:3limit}). We shall
therefore disregard this term in the sequel.  
 
Thus, \eqref{fish:ren} with $\delta F$ specified by Prop.~4.2 is the
starting point for the subsequent analysis of the boundary limit. 
In that analysis, $\delta F$ is understood as a distribution on the
differentiable functions on the interval $(-1,1)$, i.e., 
\bea
\label{Hfish}
H_\fish[f] \equiv \frac 1{i\pi} \;\delta F[f] := \int_{-1}^{+1} dv  \;
\big[h_0(v) - h_1(v)\partial_v\big] f(v).
\eea 
The crucial property will be

\medskip

\noindent {\bf Proposition 4.3:} {\sl The linear functional $H_\fish$
  vanishes on even powers $f(v)=v^{2m}$, and 
\bea
H_\fish(v^{2m+1}) = \frac 23\frac{2m+1}{2m+2}
\sum_{\nu=0}^{2m+1} J_\nu + \frac{6m+7}6 J_{2m+1} - \frac 56
\eea
where $J_n$ are given in \eqref{In}. In particular, 
$H_{\rm fish}[v^p]=0$ for $p=0,1,2,3,4$, and $H_\fish[v^5] = \frac 4{81}$.}

\medskip

\noindent {\it Proof:} The even powers of $v$ are automatically
annihilated by $H_\fish$ by symmetry under $v\leftrightarrow -v$. 
For the odd powers, see Appendix~\ref{app:limit}.

\subsection{$d=3$, $M=0$: The boundary limit }
\label{sec:3limit}

Let us first consider the most interesting case of the interaction
$\wick{\phi^4}$, i.e., $k=4$. The fish diagram contribution to the
first order correction to $(\phi^2)_{\kappa\phi^4}$ is given by  
\bea
\label{firstorder}
&& 6 \int d^3x_1 \,g(x_1)\int_0^{\infty} \frac {dz_1}{z_1^{d+1}}\;
\gamma(z_1)\,r_\fish (X_1,X) \wick{\phi^2(X_1)} =   \\
&&\qquad = \frac {6i}{8\pi^4} \int d^3y \int_0^{\infty} \frac
{dz_1}{z_1^{d+1}} \; \delta F(v)\,\theta(-y^0) \cdot
\square_{X_1}\big(\gamma(z_1)\, g(x+y) \wick{\phi^2(z_1,x+y)} \big) ,
\notag 
\eea 
where $X_1=(z_1,x_1)$, $y=x_1-x$, and $v= \frac{z^2+z_1^2-y^2}{2zz_1}$
as before. To study the boundary limit, we proceed exactly as in
Sect.\ \ref{sec:massshift}, when evaluating \eqref{Phi_1:k=2}. We
choose again $d=3$ and $M=0$. Making the same change of variables, we
put 
\bea
\Gamma(z_1,x^0-t,\vec
x+r\,\vec e_\varphi):= z_1^{-6} \;\square_{X_1}\big(\gamma(z_1)\, g(x_1)\,
(\Psi_1,\wick{\phi^2(X_1)}\Psi_2)\big) 
\eea 
and 
\bea
\Gamma_x(z_1,t,r^2):=\frac 1{2\pi}\oint d\varphi \,\Gamma(z_1,x^0-t,\vec
x+r\,\vec e_\varphi).
\eea 
Again, $\Gamma_x$ is regular at $0$, and 
\bea
\label{g0}
\Gamma_x(0,0,0)=-18\,(\Psi_1,\wick{\varphi^2(x)}\Psi_2).
\eea 
The factor $-18$ is produced by the Laplace operator \eqref{laplace}
when acting on $\wick{\phi(z_1,x_1)^2}\sim z_1^6$ at small $z_1$. Then
we arrive at the matrix element of \eqref{firstorder}
\bea
\label{firstorder2}
 = -\frac {6z}{4\pi^2}\cdot H_\fish
\Big[\int_0^\infty \!\!z_1^3\,dz_1 \int_0^\infty \!\!dt\,
\theta(t^2-w)
\cdot \Gamma_x(z_1,t,\sqrt{t^2-w})\big\vert_{w=w_{v,z}(z_1)}\Big] \quad 
\eea
which is of the same form as \eqref{HGamma}, except for
the additional power $z_1^3$ (due to the factor $\wick{\phi^2}$ in
\eqref{firstorder} as compared to $\phi$ in \eqref{log:anodim}), 
and with the functional $H$ replaced by $H_\fish$ given in \eqref{Hfish}. 

The argument in square brackets is of the form $I_3(v,z)(f)$ with
$f=\Gamma_x$ on $\RR_+^3$, as computed in the Lemma~B.1 of
App.~\ref{app:lemma}. By the same arguments as before, it
is sufficient to know it in the range $v^2<1$, where it is given by
\eqref{udvz}: there are polynomial terms $\sum_{0\leq k\leq\ell\leq 5} 
A_{k\ell}(f)\, v^k\,z^\ell$, a logarithmic contribution
\bea\label{logterms} z^5\cdot B_3(v) \cdot
\log \big((1-v)z\big) \cdot f(0,0,0)\, \qquad\hbox{with}\quad B_3(v)=
\frac {1}8\cdot v(1-v^2)(7v^2-3), \quad
\eea
and a remainder $R_{v,z}(f)=\OO(z^6)$ that 
vanishes in the boundary limit $z\searrow 0$.

By Prop.~4.3, the leading polynomial terms with $k\leq4$ are
annihilated by $H_\fish[\cdot]$, so that only the term
$H_\fish[v^5]\,A_{55}(f)\,z^5$ survives. 
The $\log(1-v)$ term in \eqref{logterms} produces another
constant
\footnote{Notice that the factor $(1-v^2)$ in $B_3$ in Lemma~B.1
    ensures the finiteness of $H_\fish[B_3(v)\log(1-v)]$.} 
times $z^5\,f(0,0,0)$, and the $\log z$-term produces the 
contribution $-\frac 78 H_\fish[v^5] \cdot z^5\log z \cdot
f(0,0,0)$. Since $H_\fish[v^5]=\frac 4{81}$ (Prop.~4.3), we
have thus found the following analog of Prop.~3.1: 

\medskip

\noindent {\bf Proposition 4.4:} {\sl For any test function $f$ on $\RR^3$,
  the limit
\bea
\lim_{z\searrow0} z^{-5} \Big\{H_\fish
\big[I_3(v,z)(f)\big] +
\frac 7{162}\,z^5\log z \cdot f(0,0,0) \Big\}\quad
\eea
is finite.} 

\medskip

Inserting this result with $f=\Gamma_x$ and \eqref{g0} into 
\eqref{firstorder2}, we find the first order contribution 
\bea
\eqref{firstorder} = -\frac {7z^6}{6\pi^2}\cdot \big(\log z\cdot
\wick{\varphi(x)^2} + \OO(z)\big)\, .
\eea
The absence of all lower order terms establishes the existence of the
boundary limit, and the presence of the logarithmic term signals the
anomalous dimension of the composite boundary field
\bea
\Delta^{\phi^2}_{\kappa\phi^2} = 6 - \frac 7{6\pi^2} \cdot 
\kappa + \OO(\kappa^2).
\eea
at first order of perturbation theory. 

{\em This establishes the existence of the boundary limit of
$(\varphi^2)_{\kappa\phi^4}$ in first order perturbation theory, when
$M=0$.} The result requires the nontrivial cancellations
$H_\fish[v]=H_\fish[v^3]=0$ of Prop.~4.3, involving the precise
functions $h_0$ and $h_1$ of Prop.~4.2 appearing in the renormalized
fish diagram. It remains to investigate whether similar cancellations
persist for $M\neq0$, $d\neq 3$, and at higher orders.   

\medskip

It is now easy to repeat the analysis for the interaction $\wick{\phi^3}$, 
i.e., $\wick{\phi^2(X_1)}\sim z_1^6$ on the r.h.s.\ of \eqref{firstorder}  
has to be replaced by $\phi(X_1)\sim z_1^3$. In this case, the power
$z_1^3$ in the $z_1$-integral is absent ($n=0$ in Lemma~B.1), hence
the logarithmic term $\log z$ arises at order $z^2$ with a coefficient
$\sim (1-v^2)$. Because $H_\fish$ annihilates the quadratic polynomial
$B_0(v)=\frac 12(1-v^2)$, but not $B_0(v)\log(1-v)$, the first-order
diagram will not contain $\log z$ terms, but finite terms $\sim
z^3\,\varphi(x)$. This reflects the expected perturbative mixing of
the fields $\phi^2$ and $\phi$ under the cubic interaction. 
Accordingly, the boundary limit $z\searrow 0$ should be taken of a
suitable combination like
$z^{-6}\big(\phi^2+\OO(\kappa)\,\phi\big)_{\kappa\phi^3}$.

\section{Conclusion}
\label{sec:conclusion}
We have pursued the strategy of perturbative construction of
interacting conformal fields in $d$ dimensions, which proceeds by the 
perturbative construction of interacting AdS fields in $d+1$
dimensions and subsequently performing a boundary limit. The
unperturbed conformal field is a generalized free field (or a
Wick product thereof). 

This procedure resolves the problematic issues associated with the
perturbation theory around generalized free fields, and at the same
time drastically reduces the expected infinite arbitrariness involved
in its renormalization. The most important benefit is the fact, that
the boundary fields, if renormalized by this method, do not suffer
from the conformal anomaly, i.e., the conformal symmetry is
perturbatively preserved.

We find, however, that the existence of the boundary limit is not
automatically guaranteed. Requiring its existence may be viewed as
another renormalization condition for the AdS field which cannot
always be fulfilled. We have pursued a number of case studies
involving polynomial interactions of scalar fields. In relevant cases,
the boundary limit exists, and the renormalized boundary fields have
anomalous dimensions that can be computed. (An anomalous dimension
does not mean a conformal anomaly!) Because the exact analytical
expressions are quite involved, we have considered only very special
cases; but in view of the highly systematic emergence of the
cancellations, we believe that the promising results found in these
cases pertain also to more general cases.  

The method is applicable only when the Lagrangean interaction
density of the conformal boundary field is induced by a polynomial
interaction on AdS. Such densities are rather special elements of the
Borchers class of the generalized free field, which carry a
reminiscence of its AdS origin. But in view of the fact that a general
perturbation theory for generalized free fields has not yet been
formulated, it is encouraging that a successful renormalization can be
achieved at least for a limited class of interactions. 

There arises an interesting question, concerning the ``continuous
operator product expansion'' for generalized free fields, as discussed
in Sect.~\ref{sec:renorm}. The OPE in the bulk is certainly a 
discrete sum. Taking the boundary limit, when it exists, should not
alter this feature. Recalling that the continuous OPE is caused
by the failure of factorization of the weight functions
$h(k_1^2,\dots,k_l^2)$ in \eqref{gwp}, we are tempted to conjecture
the perturbative stability of a discrete OPE for ``factorizing'' Wick
products whenever only the Lagrangean is a non-factorizing generalized
Wick product. To establish such a result, one would have to reorganize
the OPE of the perturbed limit fields, whose subleading terms are
continuous in terms of the unperturbed fields, into a discrete OPE in
terms of the perturbed fields.

\begin{appendix}

\section{K\"allen-Lehmann representation of 
$\Delta^{+}_{m_1}(y)\Delta^{+}_{m_2}(y)$}
\label{kallen}
\setcounter{equation}{0}
Let $\Delta^{+}_{m}(y)$ denote the 2-point function of a massive scalar
free field in $d$-dimensional Minkowski space. We are going to prove
\bea
  \Delta^{+}_{m_1}(y)\Delta^{+}_{m_2}(y)=\int_0^\infty
  dm^2\;\rho_{m_1,m_2}(m^2)\;  \Delta^{+}_m(y)
\label{KL}
\eea
with
\bea
\label{rho}
\rho_{m_1,m_2}(m^2) =\qquad && 
\\
\frac{|S^{d-2}|}
{4\cdot 2^{d-3}\cdot (2\pi)^{d-1}}&&\hskip-5mm \theta (m-m_1-m_2)\cdot m^{2-d}\Bigl(\bigl(m^2-{m_1^2}-{m_2^2}\bigr)^2-
4{m_1^2m_2^2}\Bigr)^{\frac{d-3}{2}}
\notag \eea 
where $|S^d|$ is the surface of the unit sphere $S^d$ in $d+1$
dimensions, 
\bea
  |S^d|=\frac{2\pi^{\frac{d+1}{2}}}{\Gamma (\frac{d+1}{2})}\ .
\eea

From the definitions, and using Lorentz invariance, it is easily seen,
that the K\"allen-Lehmann weight is given by  
\bea
  \rho_{m_1,m_2}(m^2)&=& \\
\frac{1}{(2\pi)^{d-1}}&&\hskip-7mm\int_{V_+}d^dp_1
\int_{V_+}d^dp_2\>\delta (p_1^2-m_1^2)\>
\delta (p_2^2-m_2^2)\>\delta (p_1+p_2-p) \notag
\eea 
where $p\in V_+$ is any four-momentum such that $p^2=m^2$. It is
convenient to choose $p=(m,\vec 0)$ and perform the integrations over
the energies $p_i^0$ first, and evaluate the momentum conservation 
$\vec p_2=-\vec p_1$. The resulting integral over $\vec p\equiv \vec
p_1$ reads in polar coordinates $p=\vert\vec p\vert$
\bea
\rho_{m_1,m_2}(m^2)&=& \\
\frac{|S_{d-2}|}{4(2\pi)^{d-1}} && \hskip-7mm \int_0^\infty\frac{dp\>\> p^{d-2}}
{\sqrt{p^2+m_1^2}\sqrt{p^2+m_2^2}}\>\delta 
(\sqrt{p^2+m_1^2}+\sqrt{p^2+m_2^2}-m). \notag
\label{rho1}
\eea 
The argument of the $\delta$-function vanishes at 
\bea
  p_0=\frac{1}{2m}\sqrt{\bigl(m^2-{m_1^2}-{m_2^2}\bigr)^2-
4{m_1^2m_2^2}}
\eea 
provided $(m^2-m_1^2-m_2^2)^2-4m_1^2m_2^2>0$ and $m-(m_1+m_2)>0$,
where the first bound is redundant. From this, we obtain \eqref{rho}.

\section{The origin of the logarithmic boundary terms}
\label{app:lemma}
\setcounter{equation}{0}
We use notations as introduced in Sect.~\ref{sec:massshift}, with
$u\equiv z_1$. For a test function $f$ on $\RR^3$, we denote by
$I(u,v,z)$ the integral  
\bea
\label{iuvz} I(u,v,z):= I(w)\big\vert_{w=w_{v,z}(u)},\quad\hbox{where}\quad
I(w):=\int_0^\infty dt\,
\theta(t^2-w)\cdot 
f(u,t,t^2-w),
\eea
and by $I_n(v,z)(f)$ the integral 
\bea
\label{invz} I_n(v,z)(f):= \int_0^\infty u^n\,du\; I(u,v,z)\qquad (n\geq 0).
\eea
We want to prove:

\medskip

\noindent {\bf Lemma B.1:} {\sl Let $z > 0$. Then $I_n(v,z)(f)$ is
  continuous w.r.t.\ $v$. In the range $v^2<1$, it is of the form 
\bea 
\label{udvz}
I_n(v,z)(f) &=& \sum_{0\leq k\leq\ell \leq n+2}
A_{k\ell}(f)\,v^k\,z^\ell  + \\ &+& B_n(v) \cdot z^{n+2} \log
\big((1-v)z\big)\cdot f(0,0,0) \;+\; R_{v,z}(f),  
\notag\eea
if $n\geq 0$ is an integer. Here, $A_{k\ell}$ are distributions,
$B_n(v)=\frac 12(1-v^2)\,{}_2F_1(-n,n+3;2;\frac{1-v}2)$ is a
polynomial of degree $n+2$, and the remainder $R_{v,z}$ is a family of
distributions that is differentiable in $v$ in the range $v^2<1$, and
that vanishes at least $\sim z^{n+3}\log z$ as $z\searrow0$. If
$n=[n]+\varepsilon$ is not an integer, then the first (polynomial) sum
extends until $[n]+2$, the logarithmic term is replaced by $C_n(v)
\cdot z^{[n]+2+\varepsilon}\cdot f(0,0,0)$ with a possibly
non-polynomial function $C_n$, and the remainder is $\OO(z^{[n]+3})$. }        

\medskip

{\bf Remark:} The emphasis is here on the various subleading terms after the
polynomial terms, because they become the leading ones in different
instances of our case studies of the boundary limit, and we expect
that this happens also in more general cases. The $\log z$-term is
essential for Prop.~3.1 and Prop.~4.4. The $\log(1-v)$-term is used in
the last paragraph of Sect.~\ref{sec:3limit}, and the
$z^{[n]+2+\varepsilon}$-term in the non-integer case is relevant in
Sect.~\ref{sec:polynomial}.  

\medskip

\noindent {\it Proof:} The integrals $I(u,v,z)$ and $I_n(v,z)(f)$ are
continuous w.r.t.\ $v$ by definition, because the integrand and the
range of integration vary continuously. For the differentiability
w.r.t.\ $v$ when $v^2<1$, we note that the dependence on $v$ is only
through $w$, and $w=w_{v,z}(u) \geq (1-v^2)u^2>0$. Thus 
$\partial_v\,I(u,v,z)=-2uz\partial_w\,I(w)$, and
\bea
\label{dwiw}
-\partial_w \, I(w) = \frac 1{2\sqrt w}\,
f(u,\sqrt w,0) + \int_{\sqrt w}^\infty dt\, \cdot 
\partial_3f(u,t,t^2-w).
\eea

We now compute the leading derivatives w.r.t.\ $z$ in the range $v^2<1$. 
Again, the dependence is only through $w=w_{v,z}(u)>0$, and
$\partial_z\,I(u,v,z)=-2(uv-z)\partial_w\,I(w)$, hence
\bea
\label{dnuiw}
\partial_z^\ell I(u,v,z) = \sum_{k=[\frac{\ell+1}2]}^\ell 
C^\ell_k\cdot
(uv-z)^{2k-\ell}(-\partial_w)^{k}I(w)\big\vert_{w=w_{v,z}(u)} 
\eea
with certain combinatorial coefficients $C^\ell_k$. Computing
$(-\partial_w)^{k}I(w)$, the derivatives can either all go on the
integrand, giving  
\bea
\label{terms1}
\int_{\sqrt w}^\infty
dt\,\partial_3^k f(u,t,t^2-w).
\eea
Or after $q<k$ derivatives on the integrand, the next derivative goes
on the lower boundary, producing $(2\sqrt w)\inv\,\partial_3^q
f(u,\sqrt w,0)$, and the remaining $k-q-1$ derivatives produce a sum
of terms (neglecting numerical coefficients for the moment) 
\bea
\label{terms2}
w^{-k+q+\frac p2+\frac12}\cdot\partial_2^p\partial_3^q
f(u,\sqrt w,0) \qquad
\hbox{with}\quad p+q\leq k-1.
\eea

Now, at $z=0$, we have $w=u^2$, hence the terms \eqref{terms1}, \eqref{terms2}
inserted into \eqref{dnuiw} become, respectively, 
\bea
\label{terms}
(uv)^{2k-\ell}\int_u^\infty dt\,\partial_3^k f(u,t,t^2-u^2),\qquad
(uv)^{2k-\ell}u^{-2k+2q+p+1}\partial_2^p\partial_3^q
f(u,u,0).
\eea
To obtain $\partial_z^\ell\,I_n(v,z)(f)$, these remain to be
integrated with $\int_0^\infty u^n\,du\ldots$. The $u$-integrals are
unproblematic at large $u$ by the falloff of the test function, but
they may become singular at $u=0$. The most singular terms are the
latter ones in \eqref{terms} when $p=q=0$, i.e.,
$v^{2k-\ell}\,u^{-\ell+1}\, f(u,u,0)$. It is then obvious that the 
$u$-integrals over \eqref{terms} are finite multiples of $v^{2k-\ell}$, 
as long as $\ell<n+2$. Thus, for $\ell<n+2$, $\partial^\ell_z 
I_n(v,z)(f)\vert_{z=0}$ is finite, and is in fact a polynomial in $v$ of
degree $\ell$, because $\big[\frac{\ell+1}2\big]\leq k\leq \ell$. 

If $n$ is an integer and $\ell=n+2$, the most singular terms
$p=q=0$ are 
\bea
\int_0^\infty du\;u^n\,(uv-z)^{2k-n-2}\,w^{-k+\frac12}\cdot 
f(u,\sqrt w,0)\big\vert_{w=w_{v,z}(u)},
\eea
with $\big[\frac{n+3}2\big]\leq k\leq n+2$. While all other terms
are finite multiples of $v^{2k-n-2}$ at $z=0$, these terms
are logarithmically divergent at $z=0$. To isolate the divergence, we
split the integration range into the intervals $(0,U)$ and
$(U,\infty)$, for any fixed $U>0$. The latter integral is a finite
multiple of $v^{2k-n-2}$ at $z=0$. In the former, we write $f(u,\sqrt
w,0) = f(0,0,0) + \big(f(u,\sqrt w,0)-f(0,0,0)\big) $, so that the
second contribution is also a finite multiple of $v^{2k-n-2}$ at $z=0$. 

The remaining terms, that diverge at $u=0$ when $z=0$, are 
\bea
\label{remaining}
f(0,0,0)\cdot \int_0^U
du\;u^n\,(uv-z)^{2k-n-2}\,w_{v,z}(u)^{-k+\frac12} =: I^{\rm div}_{n,k}(v,z).  
\eea
Restoring the suppressed numerical coefficients, these terms sum up to  
\bea
\label{sumup}
\sum_{k=[\frac{n+3}2]}^{n+2} C^{n+2}_k 
\frac 12 \Big(\frac12\Big)_{k-1}
\cdot I^{\rm div}_{n,k}(v,z) = 
 - f(0,0,0)\cdot \partial_z^{n+2} \int_0^U du\, u^n\,
 \sqrt{w_{v,z}(u)}\,.
\eea
\eqref{sumup} comprises all contributions to
$\partial^{n+2}_zI_n(v,z)(f)$ that are at divergent at $z=0$, while all
other contributions are polynomials in $v$ of degree $n+2$. 
  
The $u$-integral in \eqref{sumup} can be performed explicitly: Introducing the
integration variable $s=u-vz$ and the constant $a^2:=(1-v^2)z^2$, 
the integrand is a linear combination of terms $s^m\sqrt{s^2+a^2}$. 
If $m$ is odd, the primitive function is a polynomial in $s$, $a^2$, and 
$\sqrt{s^2+a^2}$. Evaluated at the upper and lower values $s=U-vz$ and
$s=-vz$, these are regular functions in $v$ and $z$, that possess
convergent power series expansions in $vz$ and $z$ in the range
$v^2<1, 0\leq z<U$. In particular, they contribute further finite
values at $z=0$ to \eqref{sumup}, that are polynomials in $v$ of
degree $n+2$. 

If $m=2\mu$ is even, in addition to terms of the previous algebraic
type, the primitive functions contain terms of the form 
\bea
a^{2\mu+2}\log\big(s+\sqrt{s^2+a^2}\big)\Big\vert_{s=-vz}^{U-vz} =
(z^2-v^2z^2)^{\mu+1} \log \frac{U-vz+\sqrt{w_{v,z}(U)}}{-vz+\sqrt{w_{v,z}(0)}}. 
\eea 
The logarithm of the numerator is again a convergent power series as
above, and contributes further finite values at $z=0$ to
\eqref{sumup}, that are polynomials in $v$ of degree $n+2$. 

But the denominator yields the logarithmic term
$\log\big((1-v)z\big)$. Collecting all prefactors, we find the total
logarithmic contribution to \eqref{sumup} to be given by 
\bea
(n+2)! \cdot B_n(v)\cdot \log\big((1-v)z\big)\cdot f(0,0,0) 
\eea 
with $B_n(v)= \frac 12 (1-v^2)\cdot v^n\,
{}_2F_1\big(-\frac n2,-\frac{n-1}2;2;-\frac{1-v^2}{v^2}\big)$. With
\cite[Eqs.~15.3.19, 15.3.5]{AS}, this can be brought into the
manifestly polynomial form of $B_n$ as given in the Lemma. 

Knowing (the form of) the first $n+2$ derivatives of $I_n(v,z)(f)$ at
$z=0$, we obtain the claim of the Lemma, for $n$ integer. 

If $n=[n]+\varepsilon$ is not an integer, then all terms \eqref{terms}
give rise to finite integrals $\int u^n\dots$ as long as $\ell\leq
[n]+2$, i.e., $\partial^{\ell}_zI_n(v,z)(f)\vert_{z=0}$ are
polynomials in $v$ of degree $\ell$ up to $\ell\leq [n]+2$. However, a 
scaling argument shows that $\partial^{[n]+2}_zI_n(v,z)(f)\vert_{z=0}$
has a subleading term of order $O(z^\varepsilon)$: Namely, the integrands 
\bea
g_{k}(u,v,z)=u^n\,(uv-z)^{2k-[n]-2}\,w_{v,z}(u)^{-k+\frac12}
\eea 
of the leading terms are homogeneous of order $\varepsilon-1$ in $u$
and $z$. Using Euler's equation in the form
$(z\partial_z-\varepsilon)g_{k}(u,v,z)=(-1-u\partial_u)g_{k}(u,v,z)=-\partial_u(u\,g_{k}(u,v,z))$,
this implies
\bea
(z\partial_z-\varepsilon)\int_0^U du\,g_{k}(u,v,z) = -Ug_{k}(U,v,z), 
\eea 
where $Ug_{k}(U,v,0)= v^{2k-[n]-2}U^\varepsilon$. This differential 
equation for $\int_0^U du\,g_{k}(u,v,z)$ admits
contributions $c_{k}(v)\cdot z^{\varepsilon}$ with undetermined
integration constants $c_{k}(v)$, that sum up to $C_n(v)$ in the
statement of the Lemma.

This proves the Lemma for non-integer $n$.

\medskip

The proof of the Lemma clearly exhibits the origin of the logarithmic
divergence to be the range $z_1\approx 0$ of the integration 
over $z_1\equiv u$. Notice also in \eqref{udvz} the logarithmic
singularity at $v=1$, where $\sqrt w=\vert z_1-z\vert$. It arises upon
integration over $z_1$ in the vicinity of $z$, corresponding to the
point $X_1=X$. This singularity does not lead to divergences, because
it is always tamed by the factor $1-v^2$ in $B_n(v)$.  

\section{Details of the renormalization of the 
massless fish diagram on AdS}
\label{app-pf}
\setcounter{equation}{0}
We work with the convention that the cut of $\log\, z$ ($z\in\CC$) is
along $(-\infty,0]$. As usual we define
\bea
\Li_2(z):=-\int_{C_z}dz'\,\frac{\log(1-z')}{z'}\ ,\quad
\Li_3(z):=\int_{C_z}dz'\,\frac{\Li_2(z')}{z'}\ ,\quad z\in\CC\setminus[1,\infty)\ ,
\eea
where $C_z$ is any smooth curve from $0$ to $z$ which does not
intersect $[1,\infty)$. With that $\Li_2(z)$ and $\Li_3(z)$ are
analytic on $\CC\setminus [1,\infty)$. Since
\bea
\frac{u+1}{u-1}\in (-\infty,0]\,\Leftrightarrow\, u\in [-1,1]\ ,\quad\quad
\frac{2}{1\pm u}\in [1,\infty)\,\Leftrightarrow \,u\in [-1,1]\ ,
\eea
the expression \eqref{F(u)} for $F(u)$ is manifestly analytic for
$u\not\in [-1,1]$. 

The formula \eqref{F(u)} for $F(u)$ can be derived by first computing
the integral 
\bea
F'(x)=\frac{1}{(1-x^2)^2}\int^x dt\,(1-t^2)(Q'_1(t))^2
\quad\hbox{for}\quad x\in\RR\ ,\,\, |x|>1\ ,
\eea
which gives (after analytic continuation to $z\in\CC\setminus[-1,1]$)
\bea
F'(z)=\frac{1}{(1-z^2)^2} \Bigl(\frac{2+3z-z^3}{12}\,
\Bigl(\log\frac{z+1}{z-1}\Bigr)^2 - \frac{z}3\qquad\qquad\qquad\\
+\big(\frac 16+\frac{z^2}3\big)\,\log\frac{z+1}{z-1}
+\frac 23 \,\Li_2\frac{2}{1-z}+2\,C_1\Bigr)\ ,
\notag \eea
where $C_1$ is an undetermined constant. A second integration yields
$F(u)$ for $u\not\in [-1,1]$. Here we use well-known identities for
$\Li_2$ and $\Li_3$ (see, e.g., \cite{L}) and  
\bea
\frac 1{u^2-1}\,\Bigl(\log\,\frac{u+1}{u-1}\Bigr)^{n-1}=
\frac{-1}{2n}\frac{d}{du}
\Bigl(\log\,\frac{u+1}{u-1}\Bigr)^n\ \quad (n=2,3)\,
, \\
\frac 1{1\pm u}\,\Li_2\frac{2}{1\pm u}=\mp\frac{d}{du}
\Li_3\frac{2}{1\pm u}\, .\qquad\qquad\qquad\quad
\eea
The expressions on the l.h.s.\ are problematic, since they have poles
at $u=\pm 1$, which overlap with the cut along $[-1,1]$ of the
pertinent function in the numerator. But the boundary values at
$u=v\pm i0$ along both sides of the cut of the expressions on the
r.h.s.\ are well defined distributions.

\medskip  

To compute $\delta F(v)=F(v+i0)-F(v-i0)$ we use that the complex
derivative is given by the infinitesimal differential quotient in any
direction, in particular we may choose the direction of the real axis: 
\bea
\frac{d}{dz}f(z)\big\vert_{z=v+iw}=\frac{d}{dv}f(v+iw) \quad\hbox{if $f$
  is holomorphic at $z=v+iw$,} 
\eea
and hence
\bea
\frac{d}{dz}f(z)\Big\vert^{z=v+i0}_{z=v-i0}=
\frac{d}{dv}\Bigl(f(v+i0)-f(v-i0)\Bigr)\ .
\eea
In addition we give the following formulas: 
\bea
\frac 1{u^2-1}\Big\vert^{u=v+i0}_{u=v-i0}&=&i\pi(\delta(v+1)-\delta(v-1))
=i\pi\,\frac{d}{dv}\,\theta(1-|v|)\ ,\quad v\in\RR\ ,\\
\log\,\frac{v+i0+1}{v+i0-1}&=&\log\Big\vert\frac{v+1}{v-1}\Big\vert -i\pi\,
\theta(1-|v|)\ ,\quad v\in\RR\ ,\\
\mathrm{Im}\,\Li_2(x\pm i0)&=&-\int_0^x dt\,
\frac{\mathrm{Im}\,\log(1-(t\pm i0))}{t}
=\pm\theta(x-1)\,i\pi\,\log\,x ,\;\; x\in\RR ,\qquad\quad \\
\mathrm{Re}\,\Li_2(x\pm i0)&=&\Li_2\frac{x-1}{x}+\frac 12\,(\log\ x)^2+
\frac{\pi^2}6- \log\ x\cdot\log(x-1)\ ,\quad x>1,\quad \\
\mathrm{Im}\,\Li_3(x\pm i0)&=&\int_0^x dt\,
\frac{\mathrm{Im}\,\Li_2(t\pm i0)}{t}
=\pm\theta(x-1)\,\frac{i\pi}2\,(\log\,x)^2\ ,\quad x\in\RR\ .
\eea
With that the result \eqref{dF} is obtained by a straightforward calculation 
(dropping terms involving $(1-v^2)\cdot\partial_v\theta(1-\vert v\vert) \equiv 0$).

\section{Integrals for the boundary limit}
\label{app:limit}
\setcounter{equation}{0}

Applying the functional $H_\fish[f] = \int_{-1}^{+1} dv \; 
\big[h_0(v) - h_1(v)\partial_v\big] f(v)$ (with $h_0$ and $h_1$ as in
Prop.~4.2) to odd power functions $f(v) = v^{2m+1}$, all
integrals are of the types
\bea 
J_n &=& \int_{-1}^{+1}dv\, v^n\,\log\frac{1+v}2 = (-1)^n 
\int_{-1}^{+1}dv\, v^n\,\log\frac{1-v}2, \\
K_n &=& \int_{-1}^{+1}dv\, v^n\,\log\frac{1+v}2\,\log\frac{1-v}2, \\
L_n &=& \int_{-1}^{+1}dv\, v^n\,\Li_2\frac{1+v}2 = (-1)^n 
\int_{-1}^{+1}dv\, v^n\,\Li_2\frac{1-v}2,
\eea
so that 
\bea
\label{Hodd}
H_\fish[v^{2m+1}] = \frac 23 L_{2m+1} + \frac43 J_{2m+1} -
(2m+1)\Big(\frac13 K_{2m} -\frac12 J_{2m+1}\Big) -\frac{\pi^2}9-\frac 56.
\eea
Since we could not find these integrals in the literature, we sketch
their computation here. 

In $J_n$, we partially integrate $\log\frac{1+v}2$ with primitive
$(1+v)(\log\frac{1+v}2-1)$. This gives $J_0=-2$ and the recursion
\bea J_n &=& -\frac {1+(-1)^n}{(n+1)^2}  - \frac n{n+1}\, J_{n-1}
\eea
which is solved by 
\bea
\label{In}
J_n= 2\,\frac {(-1)^{n+1}}{n+1}\,\sum_{\nu=0}^{[\frac n2]}\frac 1{2\nu+1} \, .
\eea
Summing the geometric series in the integrand of $J_n$, we also get
\bea
\label{sumIn}
\sum_{n=0}^\infty J_n= \int_{-1}^{+1}\frac{dv}{1-v}
\,\log\frac{1+v}2 = \Li_2 \frac{1-v}2\Big\vert_{v=-1}^{v=+1} =
-\frac{\pi^2}6.
\eea 

$K_n$ vanish if $n$ is odd. Partially integrating $v^{2m}$ in
$K_{2m}$, expanding $(1-v)\inv$ as a geometric series, and using
\eqref{sumIn}, we get  
\bea
\label{Kn}
K_{2m}
=  \frac 2{2m+1}\sum_{n=2m+1}^\infty J_n= 
\frac {-2}{2m+1}\Big(\frac{\pi^2}6 + \sum_{n=0}^{2m} J_n\Big)\, .
\eea

The integrals $L_n$ can be obtained by partial integration of the
factor $\Li_2\frac{1+v}2$ with primitive 
$(1-v)\big(1-\log\frac{1-v}2\big)+ (1+v)\Li_2\frac{1+v}2$, which
yields $L_0=\frac{\pi^2}3-2$ and the recursion 
\bea
(n+1)L_n = \frac{\pi^2}3 -(-1)^n\,n(J_n+J_{n-1}) - \frac{1+(-1)^n}{n+1} 
-n L_{n-1} 
\eea
with solution 
\bea
\label{Ln}
(n+1)L_{n}= \frac{\pi^2}6 -(-1)^n\sum_{\nu=n+1}^\infty J_\nu =
(1+(-1)^n)\frac{\pi^2}6 + (-1)^n\sum_{\nu=0}^{n} J_\nu.
\eea 

Inserting \eqref{In}, \eqref{Kn}, \eqref{Ln} into \eqref{Hodd} proves Prop.~4.3.

\end{appendix}

\vskip5mm

\noindent{\bf Acknowledgments.} MD profitted from discussions with
G\"unter Scharf and Raymond Stora during an early stage of this
work. Extensive discussions with Klaus Fredenhagen clarified many
conceptual issues. 
We thank the anonymous referee for insisting, by his very detailed and
qualified inquiries, on more detailed explanations in
Sect.~\ref{sec:renorm}, and for raising the interesting issue of the
structure of the OPE.

\end{document}